# Real-Time FPGA-Based CNNs for Detection, Classification, and Tracking in Autonomous Systems: State-of-the-Art Designs and Optimizations

Safa Mohammed Sali, Mahmoud Meribout, *Senior Member, IEEE* and Ashiyana Abdul Majeed

*Abstract*— This paper presents a comprehensive review of recent advances in deploying convolutional neural networks (CNNs) for object detection, classification, and tracking on Field Programmable Gate Arrays (FPGAs). With the increasing demand for real-time computer vision applications in domains such as autonomous vehicles, robotics, and surveillance, FPGAs have emerged as a powerful alternative to GPUs and ASICs due to their reconfigurability, low power consumption, and deterministic latency. We critically examine state-of-the-art FPGA implementations of CNN-based vision tasks, covering algorithmic innovations, hardware acceleration techniques, and the integration of optimization strategies like pruning, quantization, and sparsity-aware methods to maximize performance within hardware constraints. This survey also explores the landscape of modern FPGA platforms, including classical LUT-DSP based architectures, System-on-Chip (SoC) FPGAs, and Adaptive Compute Acceleration Platforms (ACAPs), comparing their capabilities in handling deep learning workloads. Furthermore, we review available software development tools such as Vitis AI, FINN, and Intel FPGA AI Suite, which significantly streamline the design and deployment of AI models on FPGAs. The paper uniquely discusses hybrid architecture that combine GPUs and FPGAs for collaborative acceleration of AI inference, addressing challenges related to energy efficiency and throughput. Additionally, we highlight hardware-software co-design practices, dataflow optimizations, and pipelined processing techniques essential for real-time inference on resource-constrained devices. Through this survey, researchers and engineers are equipped with insights to develop next-generation, power-efficient, and high-performance vision systems optimized for FPGA deployment in edge and embedded applications.

*Index Terms*— FPGA, CNN, detection, classification, tracking, Versal, quantization, DSP packing, computation model.

## I. INTRODUCTION

RECENT advancements in machine learning (ML) have transformed computer vision, particularly object detection, classification, and tracking, enabling applications in robotics, autonomous driving, video surveillance, and medical diagnostics. Object detection refers to detecting and locating instances of objects within an image, whereas classification is predicting a categorical label for a given input using patterns discovered from labeled training data. Tracking maintains object identity across frames, requiring temporal consistency and low-latency inference for real-time execution.

Early vision relied on handcrafted feature descriptors such as Histogram of Oriented Gradients (HOG), Scale-Invariant Feature Transform (SIFT), and Speeded-Up Robust Features (SURF), which lacked adaptability and scalability. [1]. The creation of Convolutional Neural Networks (CNNs) was a paradigm-shift by virtue of being able to achieve automatic, end-to-end extraction of hierarchical features directly out of raw pixels, resulting in drastic increases in both accuracy and generalization [2], [3].

Modern deep learning's compute and memory demands have driven the rise of specialized accelerators: Graphics Processing Units (GPUs), Field Programmable Gate Arrays (FPGAs), and Application-Specific Integrated Circuits (ASICs). GPUs dominate training and inference with high parallelism, but their power consumption and general-purpose nature limit edge deployments [4], [5]. ASICs offer highest performance-per-watt and small silicon footprint, suitable for latency-critical, high-volume systems (e.g., mobile, automotive) but are non-reconfigurable [5], [6], [7]. FPGAs strike a balance, offering reconfigurability, pipelined execution, custom precision, and task-specific dataflows, ideal for low-batch, streaming, edge AI [8], [9], [10].

FPGAs are ideally suited to low-batch, streaming inference applications. Latest-generation platforms such as Versal Adaptive Compute Acceleration Platforms (ACAPs), integrate reconfigurable logic with AI engines and high bandwidth memory, and enable dense and sparse execution of models [11]. Hardware-aware model optimization, such as pruning, quantization, and sparse attention, is often utilized to allow distribution of CNNs to FPGA hardware at minimal accuracy degradation [12]. Moreover, toolchains such as Vitis AI and design frameworks such as FINN [13], HLS4ML [14], and SECDA-LLM [15] accelerate CNN mapping to FPGA hardware.

FPGAs are extensively used in Advanced Driver Assistance Systems (ADAS) and edge vision. Xilinx XA Zynq UltraScale+ MPSoCs power Subaru EyeSight for adaptive cruise, lane-keeping, and collision avoidance [16]. Continental ARS540 uses the same FPGA for 4D radar processing and multi-object tracking under adverse conditions [17]. Lidar vendors also adopt FPGAs: Ouster leverages Artix-7 and Zynq UltraScale+ MPSoCs for real-time obstacle detection and point-cloud preprocessing [18], [19]. In aerospace, Northrop Grumman integrates Xilinx-based platforms like BenERA and BenPRO for high-performance satellite processing [20].

This survey reviews state-of-the-art FPGA implementations for

This work was supported by Khalifa University of Science and Technology. Safa Mohammed Sali is at Khalifa University of Science and Technology, Department of Computing and Information Engineering, Abu Dhabi, United Arab Emirates.

Mahmoud Meribout is at Khalifa University of Science and Technology, Department of Computing and Information Engineering, Abu Dhabi, United Arab Emirates (email: mahmoud.meribout@ku.ac.ae)

Ashiyana Abdul Majeed is at Khalifa University of Science and Technology, Department of Computing and Information Engineering, Abu Dhabi, United Arab Emirates (email: 100059454@ku.ac.ae).

CNN-based object detection, classification, and tracking, emphasizing hardware-aware optimizations and emerging toolchains. The main contributions are:

1) Critical review of state-of-the-art FPGA implementations of CNN-based object detection, classification, and tracking, highlighting both algorithmic innovation and hardware acceleration techniques. We also provide suggestions for improvements, taking into consideration the emergence of a new generation of FPGA processors. Potentials of integrating GPU and FPGA into hybrid architecture will also be discussed.
2) Analysis of hardware-aware optimizations such as pruning, quantization, and sparsity-aware schemes, with their accuracy and resource trade-offs, considering FPGA memory, compute, and bandwidth constraints.
3) State-of-the-art software development tools that are available for researchers to develop complex AI and non-AI applications onto advanced FPGA architectures. This aspect, which was rarely addressed in the previous survey, is important as very advanced tools are emerging to shorten the development cycle.

Table 1 highlights how our work differs from related surveys, covering recent (2025) CNN-based FPGA literature, Versal family, and sparsity handling—topics not addressed in prior CNN surveys. Fig. 1 shows publication growth (2020-2025) in real-time object detection and FPGA acceleration, underscoring the increasing importance of this domain. By bridging machine learning and hardware design, this paper serves as a comprehensive reference for researchers and system architects developing low-power, high-accuracy vision models optimized for FPGAs. Our literature coverage includes IEEE, Springer, arXiv, and US patents, identified via Google Scholar with keywords: "2025," "real-time," "FPGA," "CNN detection," "CNN classification," "CNN tracking".

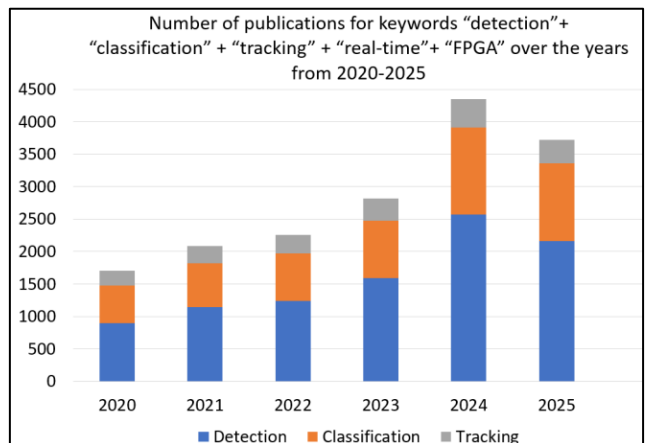

Fig. 1: Increasing trend in the field of CNN based detection, classification and tracking on FPGAs. The data is obtained from Semantic Scholar.

## II. FUNDAMENTALS AND BACKGROUND

### A. Computational Model of CNNs

AI models used for detection, classification, and tracking have similar computation models. They consist of layers which correspond to different parallelizable tasks such as convolution, normalization, etc. Table II summarizes the characteristics of these layers. In the Table, $H, W$ are the height, width of input feature map and $K$ is the kernel size. $X_i$ is feature map for channel $i$, $Y_j$ is output feature map for channel $j$. $h, w$ are the special indices in input and output feature maps. $m, n$ are kernel spatial indices. $C_{in}, C_{out}$ are the number of input and output channels respectively, while $H_{out}, W_{out}$ are the height and width of output feature maps respectively. $N_{in}, N_{out}$ are the input, output neurons, $N$ is the mini batch size (number of input samples processed together). Figure 2 graphically exhibit the meaning of these terms. All CNN models involve high-dimensional tensor operations well-suited to FPGAs, including grouped/depthwise convolutions, matrix multiplications, activations, and normalization.

TABLE I
COMPARISON BETWEEN EXISTING LITERATURES

| Year | References | Vision Tasks | Model Types | FPGA Targets | Optimization Techniques | Sparsity | Highlights |
|---|---|---|---|---|---|---|---|
| 2025 | Ours | Detection, Classification, Tracking | CNN | Xilinx Zynq, Intel, Versal, Kria | Quantization, Pruning, Pipelining. | Yes | Focuses on object detection, classification, and tracking implementation on FPGAs. Highlights sparsity and optimization. Covers hardware acceleration methods. Explores GPU–FPGA hybrid architectures for enhanced acceleration. |
| 2025 | Hozhabr etal. [21] | Detection | CNN, Transformer | Xilinx, Intel | Pruning, Knowledge Distillation (KD), Quantization | No | Focuses mainly on object detection models like CNN and transformers, HW acceleration methods, and optimization strategies to achieve real-time performance. |
| 2025 | Jiang etal. [22] | Detection | CNN | General | Pruning, KD, Quantization | No | Discussed FPGA-based hardware accelerators, optimization techniques, and compared FPGA architectures based on latency, throughput, and power. |
| 2024 | Méndez López et al. [23] | Tracking | CNN | Xilinx FPGAs, Intel Cyclone IV, V, and Intel Stratix IV GX | Pipelining, Parallelism | No | Discusses design methodologies and algorithms of visual object tracking systems on FPGAs. |
| 2024 | Bian etal. [24] | Detection | CNN | Altera's MAX series, Xilinx Zynq, Ultrascale | Pruning, KD, Quantization | No | Focuses on YOLO implementations, optimizations, and acceleration on FPGAs. |

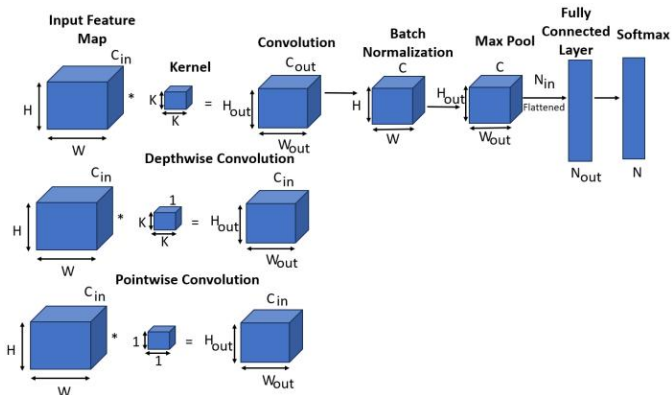

Fig. 2: Primary CNN layer types used in detection and classification pipelines. For layers following convolution (e.g., activation, normalization), C, H, and W match the output of the preceding layer.

Convolution layers dominate compute cost via dense multiply-accumulate (MAC) matrix operations across spatial and channel dimensions, while many filters and intermediate feature maps create high memory bandwidth demand. Filters (kernels) are small trainable matrices that extract features and define output depth, whereas fully connected layers perform matrix-vector operations with even higher bandwidth requirements due to dense neuron connectivity. In contrast, normalization layers like BatchNorm involve simple element-wise operations (subtract, scale, shift) with low compute cost. Pooling layers, which perform comparisons or summations to reduce spatial size, are also lightweight in computation. Activation functions add non-linearity, with ReLU being hardware-friendly and SiLU or softmax involving complex exponential/division operations. Forward propagation often exhibits irregular dependencies in multi-branch, skip-connected, or multi-scale networks, where multi-scale fusion can increase activation size to $\sum_{i=1}^{S} C_i$ channels before the next convolution layer, where $S$ is the number of scales.

Advanced versions of different CNN models have additional layers that are aimed at enhancing the model's accuracy. This includes, for instance, the self-attention or mutual attention heads mechanism, which is computationally intensive as it has $O(M^2)$ complexity for $M$ tokens. In addition, it involves a sequence of several matrix computationally intensive steps, such as linear projections, dot products for similarity scores, softmax to obtain attention weights, and weighted sum computations. To reduce the two-pass overhead of conventional SoftMax, Pillai et al. [25] describes a two-phase streaming design in their patent. During the denominator phase, exponentials of matrix-multiply outputs are accumulated in-line as they are produced, avoiding extra memory reads. A finite-state machine then triggers the numerator phase, where the buffered outputs are read once, multiplied by the stored reciprocal of the sum, and written back as probabilities, reducing latency and memory traffic by up to 50%. Table II includes some other layers which are available in advanced CNN models but expensive to handle in computationally constrained edge devices such as FPGAs. For instance, spatial pyramid pooling involves multi-pooling operations, multigrids, and feature computational complex concatenation overhead. Tiny models were suggested to some of the full advanced models to make them adapted for FPGA implementation, with acceptable draw of accuracy.

### B. Datasets for detection, tracking and classification

For computer vision models to be trained, validated, and benchmarked for tasks like object detection, classification, and tracking, datasets are crucial resources. They give models access to labelled images or videos, frequently at scale and with task-specific annotations, which facilitates fair comparison between algorithms and helps models learn visual representations.

1) **Object Detection Datasets:** Datasets like PASCAL VOC [26] provides 11,500 images across 20 classes, widely used for YOLOv1 and Faster R-CNN benchmarking. MS COCO [27] includes 118k training and 5k validation images across 80 categories with dense instance segmentation, making it a standard for modern detectors. Datasets like KITTI [28], which contains more than 7,000 annotated stereo image pairs, and BDD100K [29], which consists of 70,000 driving videos and 100,000 images labelled for object detection, present domain-specific challenges for autonomous driving applications.

2) **Object Classification Datasets:** ImageNet-1K [30], which includes 1.28 million training and 50,000 validation images spanning 1,000 object categories, is used to train CNN-based classifiers like ResNet and VGG. CIFAR-10 and CIFAR-100 [31], which contain 60,000 32x32 resolution images divided into 10 and 100 classes, respectively, are commonly used as benchmarks for lightweight models.

3) **Object Tracking Datasets:** Benchmarks like OTB-100 [32] and VOT2020 [33], which comprise brief videos with ground-truth bounding boxes and 40,000–60,000 frames, are used to assess early CNN-based trackers like SiamFC and ECO. LaSOT [34] (1,400 videos, approximately 3.5 million frames) and GOT-10k [35] (10,000 videos, approximately 1.5 million frames) support long-term, high-quality single-object tracking.

### C. Performance Metrics for Real-Time Object Detection

A wide range of performance metrics is necessary in order to impartially assess models created for object detection, classification, and tracking. These consist of energy efficiency parameters, computational complexity metrics, throughput and latency indicators, and task-specific accuracy measures.

1) **Task-Specific Accuracy Metrics:** The most commonly used metric for object detection is mean Average Precision (mAP), which provides a summary of the model's precision-recall performance across different intersection over union (IoU) thresholds. It is computed as follows:

$$mAP = \frac{1}{N}\sum_{i=1}^{N} AP_i \quad (2)$$

where $AP_i$ is the average precision for class i, and N is the number of object classes.

Multiple Object Tracking Accuracy (MOTA) and Multiple Object Tracking Precision (MOTP) are frequently used to assess tracking performance, especially in multi-object tracking (MOT). MOTA is defined as follows [36] and takes identity switches, missed targets, and false positives into account:

$$MOTA = 1 - \frac{\sum_t (FN_t + FP_t + IDSW_t)}{\sum_t GT_t} \quad (3)$$

where $FN_t$, $FP_t$, and $IDSW_t$ denote the number of false negatives, false positives, and identity switches at time t,

respectively, and $GT_t$ is the number of ground-truth objects [3]. MOTP [36] is calculated using the equation as follows:

$$MOTP = 1 - \frac{\sum_{i,t} d_t^i}{\sum_t c_t} \quad (4)$$

where $d_t$ is the distance between the localization of objects in the ground truth and the detection output and $c_t$ is the total matches made between ground truth and the detection output.

2) **Latency and Throughput:** A single input frame's processing time is known as latency. Throughput is expressed as Frames Per Second (FPS). In many applications, including autonomous vehicles, the system throughput should exceed 30 FPS.

TABLE II
COMPUTATIONAL MODELS OF CNN LAYERS

| Layer Types (Functions) | Layer Name | Primary Operation | Computational Complexity | Remarks |
|---|---|---|---|---|
| Convolutional Layers (Feature extraction) | Convolution | $Y_j(h,w) = \sum_{i=1}^{C_{in}} \sum_{m=1}^{K} \sum_{n=1}^{K} X_i(h+m, w+n) \cdot K_{j,i}(m,n)$, where $K_{j,i}$ is the convolution kernel connecting $X_i$ to $Y_j$ | $K^2 \cdot C_{in} \cdot C_{out} \cdot H_{out} \cdot W_{out}$ | Sliding-window MAC over input channels. |
| | Depthwise Convolution | $Y_i(h,w) = \sum_{m=1}^{K} \sum_{n=1}^{K} X_i(h+m, w+n) \cdot K_i(m,n)$, where $K_i$ is depthwise kernel for channel $i$ | $K^2 \cdot C_{in} \cdot H_{out} \cdot W_{out}$ | Channel-wise MAC. Used in MobileNet; reduces compute by ~9× |
| | Pointwise Convolution | $Y_j(h,w) = \sum_{i=1}^{C_{in}} X_i(h,w) \cdot K_{j,i}$, $K_{j,i} \in \mathbb{R}^{1 \times 1}$ | $C_{in} \cdot C_{out} \cdot H_{out} \cdot W_{out}$ | Element-wise linear projection-Multiply (1×1 convolution per channel combination). |
| | Transposed Convolution | $Y_j = \sum_{i=1}^{C_{in}} (X_i \uparrow) * K_{j,i}$ where $\uparrow$ denotes spatial upsampling | $K^2 \cdot C_{in} \cdot C_{out} \cdot H_{out} \cdot W_{out}$ | MAC with zero-insertion (upsampling). Used in decoders; increases spatial dimensions |
| Normalization Layers (Training stability and convergence) | Batch Normalization | $\hat{x} = \frac{x-\mu}{\sqrt{\sigma^2+\epsilon}}$, $y = \gamma \hat{x} + \beta$, where $\mu$ and $\sigma^2$ are mean and variance respectively, $\epsilon$ is small constant for numerical stability, $\gamma, \beta$ learnable affine transform params, and $x$ is input activation values to be normalized | $C \cdot H \cdot W$ | Channel-wise normalization using mean and variance. Requires division, sqrt. |
| | Layer Normalization | $\hat{x} = \frac{x-\mu}{\sqrt{\sigma^2+\epsilon}}$, $y = \gamma \hat{x} + \beta$ | $C \cdot H \cdot W$ | Normalize across features of a layer. Subtraction, division, square root, element-wise affine ops. |
| Activation Layers (Non-linearity) | ReLU | $y = max(0, x)$, where $x$ is a scalar activation input from previous layer | $C \cdot H \cdot W$ | Simple non-linearity; no multipliers needed |
| | SiLU (Swish) | $y = x \cdot \sigma(x)$, where $\sigma(x) = \frac{1}{1+e^{-x}}$ | $C \cdot H \cdot W$ | Expensive; requires sigmoid (i.e., exp/div) |
| | Leaky ReLU | $y = \begin{cases} x, & x \geq 0 \\ ax, & x < 0 \end{cases}$, where $a$ is activation value | $C \cdot H \cdot W$ | Allows small negative gradients to flow; helps avoid dead neurons in sparse activations. |
| Pooling Layers (Dimensionality reduction) | Max Pooling | $y_{i,j} = max_{(m,n) \in \mathcal{R}_{i,j}} x_{m,n}$, where $\mathcal{R}_{i,j}$ is the pooling region | $C \cdot H_{out} \cdot W_{out}$ | Maximum over local regions. Involves comparisons. Picks the most salient feature in local regions; preserves strong activations |
| | Average Pooling | $y_{i,j} = \frac{1}{|\mathcal{R}_{i,j}|} \sum_{(m,n) \in \mathcal{R}_{i,j}} x_{m,n}$, where $\mathcal{R}_{i,j}$ is the pooling region, and $|\mathcal{R}_{i,j}|$ number of elements in the region | $C \cdot H_{out} \cdot W_{out}$ | Involves add and divide. Computes mean of local features |
| Fully Connected Layers (Spatial features into class scores) | Dense Layer | $y_j = \sum_{i=1}^{N_{in}} x_i \cdot K_{j,i}, W_{j,i} + b_j$, where $x_i$ is the input neuron $i$, $y_j$ is the output neuron $j$, $W_{j,i}$ weight from input $i$ to output $j$, $b_j$ is bias term for output neuron $j$ | $N_{in} \cdot N_{out}$ | Matrix-vector multiplication |
| Output/Utility Layers (Prediction interpretation) | Softmax | $y_i = \frac{e^{x_i}}{\sum_j e^{x_j}}$ | $N$ | Normalize logits into probability distribution. Uses exponential and division. Also used as intermediate operations. |
| Skip/Utility Ops | Residual Add | $y = x + F(x)$, where $x$ is the input tensor to the residual block and $F(x)$ denotes the transformation applied through a series of layers. | $C \cdot H \cdot W$ | no computation but adds bandwidth. Element-wise addition of feature maps |
| | Concatenation | $y = concat(x_1, x_2, ..., x_n)$, where $x_n$ is the n$^{th}$ input tensor. | Sum of inputs | No computation; increases memory and layout complexity. Combine feature maps along a dimension |
| | Upsampling | $y(h', w') = x\left(\left\lfloor \frac{h'}{s} \right\rfloor, \left\lfloor \frac{w'}{s} \right\rfloor\right)$, for nearest neighbor $y(h', w') = \sum_m \sum_n x(m,n) \cdot \left(1 - \left\lvert\frac{h'}{s} - m\right\rvert\right) \cdot \left(1 - \left\lvert\frac{w'}{s} - n\right\rvert\right)$, for bilinear interpolation where s is upsampling scale factor, $h'$ and $w'$ are the coordinates in the upsampled output space | $C \cdot H_{out} \cdot W_{out}$ | Expand spatial size via interpolation or nearest-neighbor. Increases compute and memory; often used in decoders or PAN necks |

TABLE III
COMPUTATIONAL MODELS OF ADVANCED CNN LAYERS

| Layer Type | Primary Operation (Equation) | Computational Complexity |
|---|---|---|
| Self-Attention | $Y = softmax(QK^T / \sqrt{d})\,V$, where $Q = XW_Q, K = XW_K, V = XW_V$ | $O(M^2 \cdot d)$ |
| Mutual (Cross) Attention | $Y = softmax(QK^T / \sqrt{d})\,V$, where $Q = XW_Q, K = ZW_K, V = ZW_V$ | $O(M \cdot N \cdot d)$ |
| Squeeze-and-Excitation (SE) | $y_c = x_c \cdot \sigma(W_2 \cdot ReLU(W^1 \cdot z))$, $z_c = (1/HW)\sum_{h,w} x_{c,h,w}$ | $O(C^2)$ |
| CBAM (Channel + Spatial) | Channel, $M_c = \sigma(MLP(GAP(x)))$, spatial, $M_s = \sigma(f^{7\times7}([x_{avg}; x_{max}]))$ output, $y = x \cdot M_c \cdot M_s$ | $O(C^2 + C \cdot H \cdot W)$ |
| Transformer Block | $y = LayerNorm(x + MHA(x)) + FFN(x)$ | $O(M^2 \cdot d + M \cdot d^2)$ |
| Spatial Pyramid Pooling | $y = concat(pool_{n \times n}(x))\ for\ n \in \{1,2,4,\dots\}$ | $O(C \cdot H \cdot W)$ per bin |
| Inverted Residual Block | $y = x + PW_2(DWConv(ReLU(PW_1(x))))$ | $O(K^2 \cdot C + 2C^2) \cdot H \cdot W$ |

Where x,y is the input and output feature maps or tensors, $W_Q, W_K, W_V$ are learnable projection matrices for Query, Key, Value, M,N are number of tokens or spatial locations in feature maps, d is the feature dimension, $\sigma$ is sigmoid activation, FFN is feedforward network, GAP is Global Average Pooling, $f^{7\times7}$ is 7×7 convolution used in spatial attention, $K^T$ is transpose of keys, and $x_{avg}, x_{max}$ are average-pooled and max-pooled feature maps, respectively.

and $GT_t$ is the number of ground-truth objects.

3) **Computational Cost:** Model complexity is summarized by GFLOPs (floating-point ops per inference) or GMACs for low-precision arithmetic. CNN models have different requirements. For instance, while YOLO v12n requires 5.34 GFLOPs on a 640×640 input image, MobileNetV4 Conv-M used approximately 10.12GFLOps for 384×384 input image [37], [38], [39].

4) **Energy and Power Efficiency:** Energy efficiency is just as crucial for edge deployment as speed. It includes both the static power and mainly the dynamic power, which depends on the voltage and system frequency. In edge computing, usually, the power efficiency (GOPS/W) is provided as a benchmark. Hence while Xilinx Zynq7z100 FPGA yields 16.7GOPS/W, NVIDIA RTX 3090 yields 6.32 GOPS/W [132].

With the race towards incorporation of tensor compute into AI optimized hardware, the peak TOPS number is used as a key metric when comparing potential acceleration solutions. However, this can be misleading since the peak performance is only attainable when the tensor units are 100% utilized, which is usually not the case in real applications. The utilization of the tensor compute units is typically affected by two main factors: the mapping of a given workload to the available compute units, and the end to-end system-level overheads of bringing the data in/out of the chip.

## III. FPGA BASICS

FPGAs began as basic arrays of programmable logic elements in the 1980s and have since evolved into complex, heterogeneous platforms incorporating specialized accelerators and embedded processors. Over the years, this progress has led to three main architectural types: LUT-DSP based, SoC-based, and SoC-based with hardcore AI engines, which were all used in CNN for object detection, classification, and tracking, and which are described below. Two major FPGAs manufacturers have emerged for CNN-based applications: Intel (formerly Altera) and Xilinx (now owned by AMD) [40], [41] .Though Xilinx FPGAs with their advanced SOC and dedicated AI hardware in addition to advanced software development tool have a slight advantage.

### A. Classical LUT and DSP blocks FPGAs

These FPGAs are comprised of arrays of configurable logic blocks (CLBs), each of which has multiplexers, flip-flops, and look-up tables (LUTs). These blocks are connected by a flexible routing fabric [42]. A K-input LUT in a CLB stores output values in SRAM (or flash) cells, allowing it to implement any Boolean function of up to K variables [43]. This structure allows for fine-grained, bit-level datapath customization, enabling designers to implement arbitrary quantization schemes including INT2, INT4, INT6, INT8, INT16, and even non-uniform formats. This permits bit-accurate control but has a lower throughput and higher resource consumption than devices with dedicated arithmetic engines. LUT-DSP based FPGAs like the Lattice iCE40, Xilinx Spartan-7, and Intel Straix 10 provide the most flexibility in applications where speed and resource efficiency are not critical, such as testing new quantization schemes, developing energy-efficient microcontrollers, or prototyping quantization-aware models. However, larger bit-widths like INT16 or FP16 are slower and area-inefficient because all arithmetic must be manually mapped into logic. They also comprise an array of DSP and RAM blocks to implement fast MAC of up to two 54 x 54 operands.

Lut-DSP based FPGAs are ideal for real-time applications that demand deterministic latency and predictable timing closure. Among these, the Artix UltraScale+ XA AU7P FPGA [44] stands out as a compact 9×9 mm automotive-grade FPGA built on 16 nm FinFET technology, specifically designed for high-reliability vision workloads in harsh environments. Its column-and-grid UltraScale+ architecture organizes CLBs, DSPs, BRAM, I/Os, and transceivers in uniform clock regions with low-skew routing, simplifying timing closure and ensuring microsecond-level response.

The AU7P integrates 6-input LUT CLBs with 216 DSP48E2 slices, each a 27×18 multiplier with 48-bit accumulator, supporting SIMD modes (dual 24-bit or quad 12-bit) for INT8/INT16 CNNs and DSP tasks. Floating-point (FP16/FP32) can be synthesized via Vivado Floating-Point IP, trading area and power for mixed-precision flexibility. This balanced design enables low-latency, resource-efficient CNN inference under strict automotive condition. Fig. 3 depicts the top-level arrangement of functional resource blocks within the device.

In LUT DSP FPGAs, AI accelerators are built from Processing Elements (PEs), which are small compute blocks with local BRAM/registers and MAC units (2 FLOPs per MAC). When tiled into a 2D systolic array (regular PE grid where inputs flow vertically, weights horizontally, and partial sums diagonally), this structure achieves high spatial parallelism and data reuse [45]. Performance is enhanced by loop unrolling (replicating loops to expose parallelism) [46], tiling (splitting large feature maps to fit

BRAM), and pipelining with initiation interval (II) = 1, ensuring a new input is processed every cycle. Double-buffered feature maps sustain throughput, while MACs map to DSP slices and weights/activations to BRAM, forming the backbone of CNN acceleration on LUT-DSP FPGAs one [47], [48], [49].

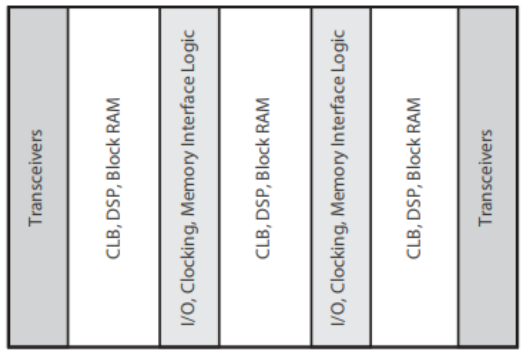

Fig. 3: FPGA with Columnar Resources from [44]

### B. System on Chip (SoC) FPGAs

Modern SoC FPGAs from Xilinx and Intel integrate hard ARM cores with the programmable logic (PL) to form a processor system (PS) that offloads sequential tasks to CPUs while the PL accelerates parallel workloads. The PS, known as the Application Processing Unit (APU) in Xilinx and Hard Processor System (HPS) in Intel, comprises Cortex-A cores, peripherals, memory controllers, and DMA engines, and communicates with the PL via AXI-GP (control), AXI-HP (high-throughput), and AXI-ACP (cache-coherent) interfaces. Representative devices include Intel Agilex SoC [50], Zynq UltraScale+ MPSoC [51], and Xilinx Zynq 7000 [41]. The PS supports LPDDR3/4 and DDR4/5 with up to 230.4 Gbps bandwidth (72-bit DDR4-3200 in Agilex F-Series) and often integrates HDMI and H.264 decoders, enabling edge-AI video applications. Within the PL, Xilinx DSP48E1/E2 slices (25×18 multiplier, pre-adder, 48-bit accumulator, SIMD dual-24/quad-12) deliver INT8/INT16 MACs with optional FP16/BF16, while Intel Agilex DSPs (up to 54×54 multipliers with ALMs and M20K RAMs) support INT4-INT16, FP16, and FP32 [52].

Although FPGA DSP blocks are optimized for standard widths (mainly INT8 for ML inference), designers can implement custom or mixed-precision arithmetic (e.g., INT5, INT7) in LUTs, at the cost of higher CNN latency due to less-optimized routing. Toolchains are critical: Xilinx's FINN [53] compiles quantization-aware networks to PL with arbitrary precisions down to 1 bit, while Xilinx's Vitis AI [54] supports only INT8 models, deploying them on the soft DPU (DPUCZDX8G) [54], a LUT-DSP-based systolic array, as shown in Fig. 4. The DPU communicates with the APU via AXI interconnects; its Instruction Fetch Unit sends commands to a High-Performance Scheduler, which performs static compile-time scheduling of CNN layers onto a Hybrid Computing Array. Hybrid Computing Array is a grid of PEs optimized for parallel MAC operations. Data is streamed from external DDR through a High-Speed Data Pipe into the Global Memory Pool, a shared on-chip buffer that stores feature maps, weights, and intermediate results to feed the computing array. This layer-fusion execution (e.g., Conv + ReLU) minimizes intermediate memory accesses and supports continuous PE operation. The DPUCZDX8G is offered in multiple configurations, with the B4096 variant achieving 4,096 INT8 MACs per cycle for convolution and fully connected layers.

The DPU, implemented entirely in programmable logic, achieves high CNN parallelism but incurs higher logic utilization and tens of cycles of latency per layer, depending on network depth and memory access. Integrated with Vitis AI, it simplifies platform-portable CNN deployment without requiring HDL expertise. The capability of SoC FPGAs to carry out real-time task scheduling within the PS is an additional strength. ARM cores are used by real-time operating systems (RTOS) like FreeRTOS or embedded Linux with PREEMPT-RT patches, which allow for deterministic scheduling of time-sensitive software tasks, interrupt management, and peripheral control [55]. However, the RTOS aspect of effectively handling PS-PL interaction in SOC-FPGA was not tackled in the literature, including for AI-based video applications. This can be an attractive research venue, especially to implement multiple AI models with different levels of priorities and worst-case scenario delays (WSCT).

Intel's Agilex 3 FPGA architecture [50], fabricated on Intel 7 with second-generation HyperFlex, exemplifies advanced SoC FPGAs, as illustrated in Fig. 5. It adopts a registers-everywhere architecture that includes Hyper-Registers (bypassable flip-flops along routing paths for fine-grained timing), Hyper-Retiming (relocation of these registers to equalize logic delays), and Hyper-Pipelining (insertion of additional pipeline stages without RTL changes). Together, these features enable >1 GHz clocking and support deep pipelining. The PL integrates 34k ALMs, 262 M20K RAMs, and 276 variable-precision DSPs in a columnar grid, while AI Tensor Blocks support INT8, INT16, and mixed-precision (e.g., INT4 early, INT8 later) to balance accuracy and efficiency.

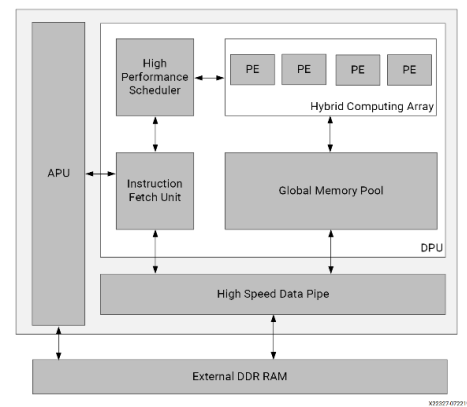

Fig. 4: Top level architecture of DPUCZDX8G [56] .

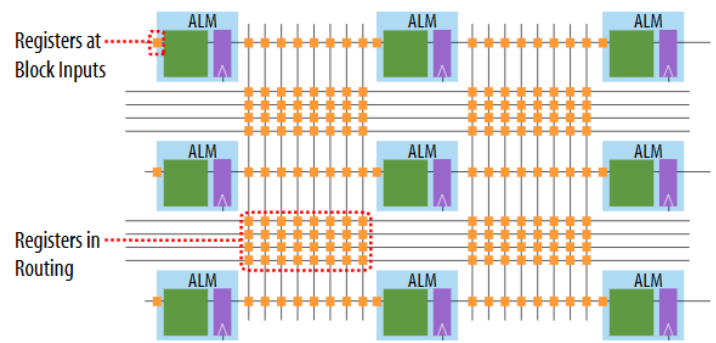

Fig. 5: Hyperflex's "Registers Everywhere" architecture. The orange boxes correspond to the Hyper-Registers in the core fabric [57].

Compared with Xilinx devices, Agilex 3 offers superior timing closure and pipelining flexibility, with its dual-core Cortex-A55 MPCores and FP32/FP64 support enabling hybrid AI workloads.

## C. SoC FPGA with hardcore AI engines

Versal devices extend the PS–PL continuum with a third heterogeneous tier: a two-dimensional mesh of hard AI Engine tiles (AIE and AIE-ML), each incorporating a SIMD very-long instruction word (VLIW) core and 128-bit AXI-Stream NoC routers [58], [59]. These VLIW-based cores can execute multiple operations per cycle across scalar, vector, and address-generation units. The vector registers natively support a wide range of data types including int8/uint8, int16/uint16, cint16, int32/uint32, cint32, BF16, FP16, and int64/uint64, with grouping sizes of 128, 256, 512, or 1024 bits for AI kernels implemented via the AI Engine API (eg., Versal AI Core series, Versal AI Edge). The total data width of the vector units can thus scale up to 1024 bits per cycle, offering high parallelism for both integer and floating-point workloads.

AIE tiles in Versal AI Core/Premium provide 32 KB SRAM, while AIE-ML tiles in AI Edge/HBM and 2nd-gen AI Core offer 64 KB SRAM with integrated DMA for efficient tensor movement. All engines share a multi-Tb/s NoC connected to DDR4/LPDDR4 controllers, PCIe Gen4/5, and GTY/GTYP transceivers, supporting layer-distributed systolic execution and multi-FPGA scaling. As shown in Fig. 6b, each tile integrates an AI Engine, tile interconnect, and memory module; the data memory is divided into eight banks with DMA and locking for synchronized multi-tile access. The scalar processor, 512-bit SIMD vector processor, three address generators, and 16 KB program memory work with a cascade-stream interface, allowing accumulator outputs to flow to neighbouring tiles for pipelined multi-layer CNN execution. Fig. 6a illustrates the cascade flow: data moves horizontally bottom-to-top across tiles, then vertically at row ends, alternating east–west directions for a continuous pipeline. Non-adjacent communication uses programmable AXI4-Stream interconnects for scalable inter-tile routing.

The AIE MAC pipelines, together with Versal soft DPU, DPUCV2DX8G [60] which is integrated with systolic architecture, are optimized for asymmetric per-tensor INT8 weights and activations. On a fully populated VC1902, this combination can exceed 100 TOPS. A compile-time scheduler orchestrates both AI-Engine tiles and the hard DPU while shared-weights buffer reuses parameters and slash external-memory bandwidth. A low-latency SHIM plus on-chip NoC fabric link AI Engines, the DPU, PL, and off-chip DDR, together delivering fused, high-throughput CNN inference for vision workloads, as illustrated in Fig. 7.

The Versal AI Core series lacks high-bandwidth memory (HBM), whereas Versal HBM devices (e.g., XCVH1582) integrate HBM2e via a 1024-bit interposer interface, achieving up to 820 GB/s bandwidth for memory-intensive AI workloads [61]. There are no Intel devices with AI engines similar to Versal.

Following Table IV's comparison based on LUTs, DSPs, presence of embedded processors, on-chip/external memory, frequency and quantization flexibility of the three types of FPGAs, the choice among these architectures for CNN inference is driven by application priorities. The most cost-sensitive deployments benefit from LUT-DSP FPGAs, whose fine-grained logic fabric supports arbitrary quantization 1–16 bits) and delivers modest throughput ($\approx$1–143 TOPS) at low efficiency ($\approx$0.2–1 TOPS/W), while guaranteeing deterministic, low-latency pipelines for ultra-light models or energy-constrained vision nodes. SoC-based FPGAs offer a balanced solution when integration and moderate performance are required. They offer 2–8 TOPS at 1–4 TOPS/W, making them well suited to smart surveillance, drones, and robotics with CPU-driven pre- and post-processing. Finally, AI-augmented ACAPs (e.g., AMD Versal) achieve maximum throughput and sparsity-aware acceleration via hard AI Engines, optional HBM2e (>800 GB/s), and mixed-precision support (INT8/BF16/FP16), exceeding 100 TOPS at 8–12 TOPS/W. Despite higher cost, these devices are the preferred choice for real-time, large-scale inference tasks where both performance and efficiency are paramount.

Table V highlights commonly used FPGA boards for real-time vision tasks, focusing on compute, memory, and cost. The AMD ZCU104 offers balanced performance with 230K LUTs and 1,728 DSPs, priced around $1678. The Versal VC1902 delivers high-end capability with 400 AI Engines and nearly 2K DSPs, but costs around $13,000, making it suitable for demanding AI tasks. For lower-power deployments, the Kria KV260 is a compact and affordable option with 70K LUTs. Intel Agilex 7 offers high DSP and memory capacity, while Arria 10 GX provides a mid-range choice for moderate CNN workloads. These options offer trade-offs between performance and cost depending on application needs.

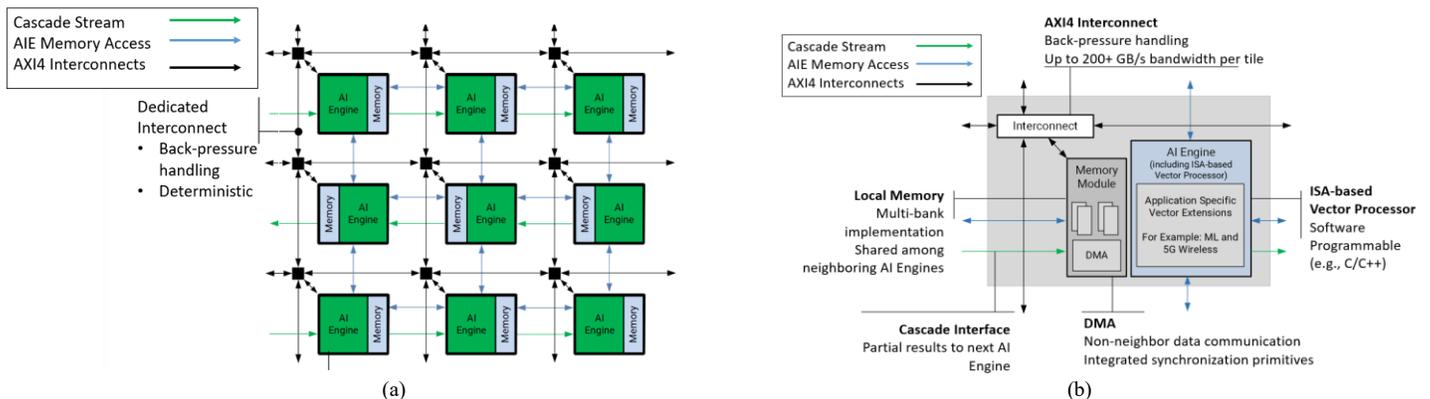

Fig. 6: (a) AI Engine Array (b) AI Engine Tile from [58]. The AI Engine can access memory modules in all four directions, treating them as a contiguous block, with the memory interface directing each access to the appropriate direction based on the generated address.

TABLE IV
COMPARISON BETWEEN LUT-BASED, SOC LUT-BASED, AND ACAPS

| FPGA Classifications | LUTs, DSP Slices | Embedded Processors | On-chip SRAM (BRAM / URAM / NoC SRAM) | External DRAM & peak bandwidth | Quantization Support | Typical max clk (fabric / CPU-class core) | INT8 TOPS / TOPS per W |
|---|---|---|---|---|---|---|---|
| LUT-DSP-Based FPGAs | ~10 k – 3 M, 48–12300 DSP48-class slices | None | up to ≈ 455 Mb BRAM + URAM | Soft-IP support for DDR/DDR2/DDR3/DDR4/QDRII+/RLDRAMII/III typical BW < 8 GB/s per channel | Any precision (1–16 bits, custom fixed-point) | 50 – 200 MHz | ~1 – 143 TOPS / ~0.2 – 1 TOPS/W |
| SoC-Based FPGAs | 50K – 2M+ LUTs, ~2 000 – 5 000 DSPs | Hard ARM Cortex-A9/A53 cores (up to quad-core) | 5 – 66 Mb BRAM + ~256 KB OCM | DDR3/4/5, LPDDR4, up to ~1.1 TB/s | INT8, INT16, INT18, INT27 in DSPs (Xilinx), INT4–INT27, FP16, FP32 for Intel devices | 200–400 MHz (PL) / 0.5–1.3 GHz (PS) | ~2 – 8 TOPS / 1 – 4 TOPS/W |
| SoC FPGA with hardcore AI engines | 100K – 2M+ LUTs, Up to ~7K (ACAP total) + AI Engine MACs (up to 400 AIE tiles) | Dual Cortex-A72, dual Cortex-R5, AI Engines (SIMD VLIW cores) | 40 – 245 Mb total on-chip SRAM + AI Engine memory | LPDDR4 / DDR4 (32-64 bit) In-package HBM2e: 32 GB, up to 820–880 GB/s (HBM series) | INT8/INT16/INT32, BF16, FP16, CINT16, CINT32 with AI Engines and DSP58, DSPFP32 and DSPCPLX Modes | 400–600 MHz (PL) / 1.3 GHz (AI Eng.) / 1.5 GHz (A72) | INT8: > 100 TOPS; FP16/BF16: ~ 50 TOPS; TOPS/W ≈ 8–12 |

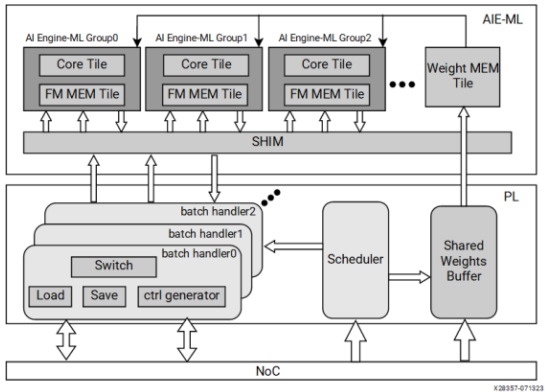

Fig. 7: Block diagram of DPUCV2DX8G [60].

TABLE V
COMMONLY USED FPGAS FOR REAL-TIME VISION TASKS

| FPGA Board | LUTs / ALMs, FFs | DSP /AI engine | Block RAM (size) | Cost |
|---|---|---|---|---|
| AMD Xilinx Zynq UltraScale+ ZCU104 [62] | 230,400 LUTs, 461,000 FFs | 1,728 | 146 (38Mb) | $1,678 |
| AMD Versal AI Core VC1902 [63] | 899,840 LUTs, 1,799,680 FFs | 1,968 DSPs 400 AI | 34Mb | $13,195 |
| AMD Xilinx Kria KV260 [64] | 70,560 LUTs, 141,120 FFs | 1.2K | 144 | $249 |
| Intel Agilex 7 AGM 032 [65] | 1,100,000 ALMs, 440,0000 FFs | 9,375 | 15,932 (311Mb) | $9,495 |
| Intel Arria 10 GX 1150 [66] | 427,200 ALMs, 1708800 FFs | 1518 | 2,713 (52.96Mb) | $10,019 |

Recent edge AI platforms show a trade-off between compute performance and efficiency. AMD's Versal AI Edge Gen 2 [67] delivers up to 185 dense INT8 TOPS, 370 sparse TOPS, and ~12–15 TOPS/W, offering reconfigurability and sparsity-aware acceleration. NVIDIA's Jetson Thor [68] achieves 2070 TOPS at 15.9 TOPS/W, leading in raw performance. Google's Coral M.2 Accelerator with Dual Edge TPU [69] provides 8 TOPS at 2 TOPS/W, targeting low-power, embedded applications. Versal balances efficiency and flexibility, while Jetson favors throughput and the TPU prioritizes minimal power.

## IV. FPGA-BASED CNN IMPLEMENTATION

To achieve real-time inference, CNN architectures frequently require gigabytes of memory, teraflops of compute performance, and extensive parallelism. Meeting such demands becomes especially difficult when deployed in embedded or edge scenarios because of limitations on hardware resources, power, and latency. Due to their low execution latency, reconfigurability, and adaptable parallel datapaths, FPGAs have become a viable option for speeding up deep learning workloads. But translating these intricate models to FPGA platforms necessitates algorithmic simplification, memory optimization, and careful hardware-software co-design. In this section, we examine parallelization and optimization techniques for CNN tasks on FPGAs.

*A. CNN Parallelism and FPGA Resource Optimization*

CNNs are well-suited for FPGA acceleration due to their spatially shared weights, localized receptive fields, and regular compute patterns. These properties enable high data reuse and predictable memory access, aligning perfectly with the fine-grained parallelism of FPGAs. For instance, a single 3×3 convolution with 64 input and output channels on a 224×224 feature map involves over 186 million MACs [70].

Multiple axes of parallelism are utilized by FPGA accelerators. Kernel-level parallelism involves unrolling convolution loops over the kernel's spatial dimensions. For example, all the multiplications within a 3×3 kernel can be executed simultaneously, increasing throughput for each convolution operation. Spatial parallelism computes multiple output pixels at once by parallelizing across the spatial dimensions of the feature map. Channel-level parallelism processes multiple input or output channels concurrently. However, under constrained FPGA resources, each kind of parallelism has trade-offs. For instance, despite providing superior DSP efficiency, channel-level parallelism necessitates sizable on-chip buffers to store numerous activation and weight channels. In contrast, when feature map sizes are small, spatial unrolling may underuse DSPs but can result in better PE utilization. The extent of parallelism directly impacts the resource footprint defined by the usage of LUTs, DSPs, BRAMs, and routing. Although CNNs have a nonlinear resource scaling, they are structurally ideal for FPGA acceleration.

Fig. 8 shows an FPGA-based systolic CNN accelerator that exemplifies many of these principles, including PE-level tiling, local buffering, and minimized off-chip access [71]. By distributing MAC operations across a 2-D array of PEs and overlapping data fetch, sparse matching, and accumulation in pipelined buffers, the design achieves high spatial parallelism. Activations are fetched from DRAM into the Input Buffer, while sparse weight tiles and indices are loaded through the Weight Buffer and Weight Index Buffer. The Input Matching Unit (IMU) aligns activations and weights based on their sparse indices before feeding them into a 2-D systolic PE array. Final activations and pooling are performed using a tree-adder, Output Buffer, and Vector Processing Unit (VPU). This design efficiently integrates tiling, sparse computation, and on-chip buffering to minimize off-chip memory access and maximize compute throughput. This sparse-periodic systolic dataflow on VGG16 architecture with Xilinx VU9P FPGA consumed only 15% of DSPs, 10% of LUTs and 12% of BRAMs.

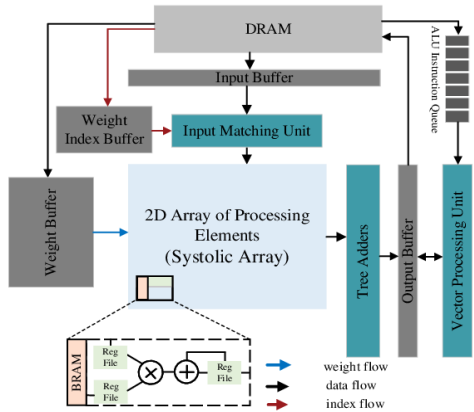

Fig. 8. An overview of the sparse -periodic systolic CNN accelerator from [71].

### B. Optimization Techniques in FPGAs

Exploiting the spatial parallelism of FPGAs under strict BRAM, DSP, and bandwidth constraints demands careful algorithm-hardware co-design. This section outlines key optimization strategies that make real-time deep learning models viable for real-time FPGA deployment.

#### 1. Spatial Dataflow & Pipelining

Spatial dataflow pipelining sustains continuous operation by overlapping computation and data movement, allowing new inputs every clock cycle while reusing intermediate activations on-chip to reduce DRAM access [72]. Critical techniques include layer fusion, balanced pipeline stage partitioning, and shared arithmetic resource reuse, all crucial for deep CNNs and detection networks under latency constraints. Different types of parallelism techniques are illustrated in the table VI.

TABLE VI
TYPES OF PIPELINING

| FPGA Board | LUTs / ALMs |
|---|---|
| Layer-level pipelining | Overlaps the execution of different layers for the same input by streaming intermediate outputs directly from one hardware stage to the next without writing back to memory. |
| Intra-layer pipelining/unrolling | Unrolls and pipelines operations within a single layer (e.g., convolutions across pixels/channels) so that each new computation begins every clock cycle (II = 1). |
| Layer fusion | Combines multiple operations like conv + BN + ReLU into a single hardware block, reducing memory usage and latency between layers. |

Fig. 9 shows H2PIPE, a high-throughput spatial pipeline [73]. where convolution, depthwise, and pointwise layers are mapped to dedicated hardware modules connected by activation buffers in an assembly-line dataflow. On Intel Stratix 10 NX, H2PIPE achieves >1000 FPS on ResNet-50 and <1.1 ms latency on ResNet-18, demonstrating that spatial mapping and careful pipelining enable real-time inference for large models.

However, these gains come with trade-offs. Layer-level pipelining demands large activation buffers between stages, which can significantly increase BRAM usage and overall latency if not properly tuned. Variations in layer complexity also make pipeline balancing difficult, potentially leading to stalling or underutilization of hardware resources. Intra-layer optimizations such as loop unrolling exploit fine-grained parallelism but can quickly exhaust DSPs and routing resources, especially for wide channels or large feature maps. While layer fusion helps reduce latency and off-chip memory traffic, it also introduces control complexity and reduces design modularity, complicating debugging and hardware reuse. These challenges are critical on resource-constrained FPGAs, making automated design-space exploration with resource-aware performance models essential.

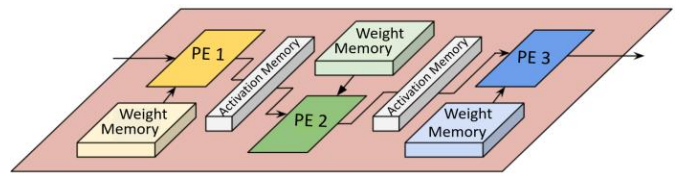

Fig. 9. Overview of CNN dataflow architecture showing layer-level pipelining. Each PE has its own weight memory. Activation FIFOs between PEs enable continuous streaming of data every clock [73].

#### 2. Precision Scaling & Quantization

Precision scaling is critical for real-time FPGA inference, reducing weights, activations, and intermediate signals from FP32 to low-precision fixed-point formats (INT8, INT4, binary) to cut memory bandwidth and resource usage. Lower precision enables more parallel PEs on the FPGA fabric, improving throughput and energy efficiency for latency-sensitive tasks such as object tracking and detection. Unlike GPUs, FPGAs offer flexible logic that efficiently supports low-bit operations, improving throughput per watt. For example, on Xilinx UltraScale+, DSP48E2 slices can pack up to four INT4 MACs per DSP, greatly boosting compute density and energy efficiency versus INT16 execution. These capabilities drive precision-scaled accelerators that pair Quantization-Aware Training (QAT) with optimized computation to achieve high-speed inference without accuracy loss.

Post-Training Quantization (PTQ) and QAT are two examples of contemporary quantization techniques. PTQ statistically calibrates fixed-point scales (e.g., percentile clipping) but may lose accuracy below 8-bit, while QAT simulates quantization noise during training using fake nodes, enabling adaptation to low-bit formats (INT4, INT2) and preserving accuracy. Uniform quantization with fixed step sizes is best suited to FPGA hardware, whereas non-uniform or block-floating formats expand dynamic range but require extra LUTs and control logic. In practice, Peccia et al. [74] demonstrated PTQ on YOLOv7-tiny, converting it to INT8 for a Gemmini-based FPGA accelerator, preserving detection accuracy by skipping Non-Maximum Suppression (NMS) quantization and using per-tensor scaling for a balance of simplicity and precision.

Despite these gains, quantization presents notable challenges. Aggressive precision reduction can significantly degrade accuracy, particularly in early layers and residual paths. While QAT can mitigate this, it introduces additional training complexity and tuning overhead. Ultra-low precisions (binary/INT2) often fall back to bit-serial or LUT arithmetic, reducing DSP utilization and increasing latency. Mixed-precision designs improve flexibility but may fragment resources due to DSP/BRAM granularity. Future progress relies on quantization-friendly networks, bit-accurate simulation, and compiler automation to enable workload-aware precision mapping and efficient FPGA utilization.

### 3. DSP Packing and Low-Bitwidth Arithmetic Optimization

DSP packing leverages low-bitwidth quantization (e.g., INT4/INT8) to execute multiple MACs per DSP slice, boosting compute density. Langhammer et al. [75] demonstrated an architecture in their patent with multiple weight-register columns and a bank of parallel multipliers, enabling each weight to be multiplied concurrently with all elements of an input vector, effectively doubling or more the MACs per cycle without extra DSP slices. This approach maximizes multiplier utilization and reduces off-chip memory traffic for CNN inference.

On Xilinx Zynq UltraScale+, 1:1 mapping of INT8 MACs to DSP48E2 slices underutilizes the full multiplier width, limiting compute density. Peccia et al. [74] addressed this by packing two INT8 MACs per DSP48E2 using operand partitioning. This doubled arithmetic density without extra DSPs, allowing the systolic array to scale from 16×16 to 32×32 PEs, while DSP usage rose only from 441 to 652, far below the 4× growth expected, yielding a 60% mean speed-up over the baseline Gemmini.

While packing is highly effective for fixed-precision (INT8) networks, it is less suitable for binary or mixed-precision workloads, which may require bit-serial logic or alternative arithmetic paths. Bit alignment and masking are essential to prevent numerical errors, adding design and verification overhead. Additionally, non-MAC operations (e.g., activation, normalization, residuals) often cannot map efficiently to DSPs and must fall back to general logic or CPUs, complicating memory scheduling and resource allocation, particularly for multi-model deployments.

Despite these limitations, DSP packing remains a key FPGA optimization for CNN inference, offering high compute density, power efficiency, and deterministic latency. Combined with quantization and systolic dataflows, it enables high-throughput designs under tight resource budgets.

### 4. Pruning & Sparsity

Pruning reduces CNN complexity by removing redundant weights, thereby lowering compute cost and memory access. This introduces sparsity, the condition where many weights or activations are zero, enabling optimizations like zero-skipping. Unstructured pruning zeroes individual weights, achieving high sparsity but causing irregular memory access that degrades FPGA performance. Structured pruning, which removes entire filters, channels, or blocks, aligns better with FPGA spatial parallelism and regular memory access, supporting predictable high-throughput inference for tasks such as classification and object detection.

Fig.10 illustrates the pruning-aware FPGA accelerator by Peccia et al. [76] which employs a Hardware Aware Pruning Method (HAPM). HAPM groups weights to match the systolic array's parallel kernel mapping and prunes them iteratively by magnitude to maintain compatible sparsity patterns. A Dynamic Sparsity Bypass (DSB) skips zero-valued MACs directly within input buffers, reducing unnecessary computation while keeping the array simple and throughput-saturated. Applied to ResNet on CIFAR-10, this approach reduced latency by 45% with 2.5% accuracy loss, achieving 7.468 GOPS using 144 DSPs on a ZedBoard.

Pavlitska et al. [77] addressed structured pruning challenges in architectures like YOLOv7 by proposing a connectivity-aware graph pruning method that maintains channel dimensions across concatenation paths. As shown in Fig.11, accuracy declined with increasing sparsity, and model sparsity alone did not guarantee speed-up.

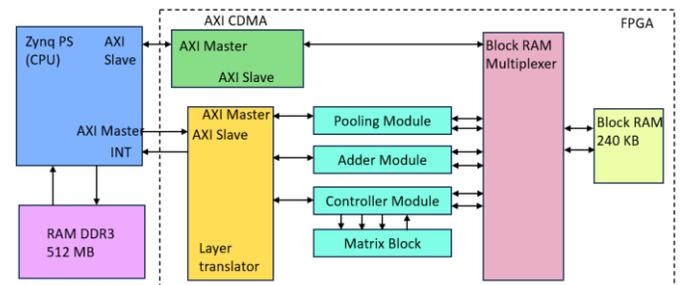

Fig. 10. FPGA-based CNN accelerator from [76], integrating a Zynq SoC, DDR3 with CDMA, and a modular pipeline of controller, adder, pooling, and systolic-array matrix blocks coordinated by a Layer Translator for structured data reuse and DSB.

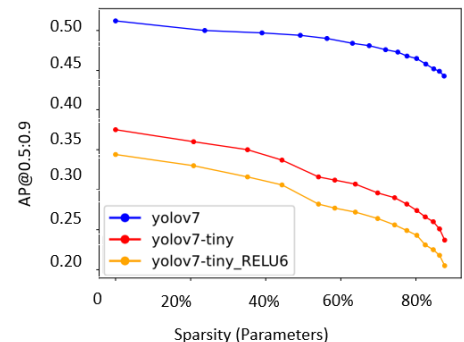

Fig. 11. Accuracy for different levels of sparsity measured in terms of parameters [77].

## V. Software Toolchains for Real-Time Detection, Classification, and Tracking on FPGAs

To meet the strict latency, throughput, and power constraints of embedded vision systems, real-time object detection, classification, and tracking on FPGAs necessitate a tightly integrated suite of software tools.

### A. Xilinx Toolchain (Vitis Ecosystem)

Xilinx's software ecosystem, offers a tightly integrated flow that simplifies hardware/software co-design and supports end-to-end deployment of AI models on edge platforms. The key advantage lies in its mature support for HLS, a wide range of DPU overlays, seamless integration with popular AI frameworks, and profiling/debugging tools. Additionally, Docker-based deployment and Python APIs ease rapid prototyping. The following are the main tools used for CNN vision tasks.

1) **Vitis unified software:** This software combines High-Level Synthesis (HLS) accelerators with C/C++ applications for Zynq UltraScale+ and Kria SOMs and enables hardware/software co-design [78].
2) **Vivado Design Suit:** It offers low-level support for FPGA synthesis, AXI interconnect management, and IP integration [79].
3) **Vitis HLS:** It implements real-time modules (e.g., feature extraction, correlation filters) using loop unrolling, pipelining, and interface pragmas [80].
4) **Vitis Video Analytics SDK (VVAS):** Provides a GStreamer-based framework for real-time video analytics, integrating hardware inference, overlay rendering, and camera capture into a single pipeline that can be fully offloaded to FPGA fabric when necessary [81].
5) **Vitis AI:** It is a primary tool for deploying AI models on FPGAs [82], streamlining model training to hardware inference for TensorFlow, PyTorch, and Caffe. As shown in Fig.12, the stack begins with the AI Model Zoo of pre-trained vision models, which are quantized from FP32 to INT8 via the Vitis AI Quantizer using calibration and optional fine-tuning. The Vitis AI Compiler converts these into .xmodel executables, mapping layers to DPU resources. VART (Vitis AI Runtime) provides C++/Python APIs for model execution, buffer management, and DPU interfacing, while Xilinx Runtime (XRT) handles low-level scheduling and PS-PL data transfers. Vitis AI Libraries offer pre-optimized pre/post-processing kernels, accelerating end-to-end development. Vitis AI Profiler enables layer-wise latency and throughput analysis, supporting iterative real-time optimization. The framework also supports Docker-based deployment and cross-compilation for Zynq UltraScale+ MPSoC, Versal ACAP, and Alveo cards.

The major drawback of Vitis AI is that it can only accept FP32 for converting to INT8 during quantization which create constraints during training. Moreover, the reliance on predefined DPU overlays may limit adaptability for custom or irregular network topologies.

### B. Intel toolchain

Intel FPGAs use the OpenVINO toolkit [83] to accelerate edge AI. As shown in Fig. 13, models trained in TensorFlow, Keras, or ONNX are first converted to an Intermediate Representation (IR) using the Model Optimizer. The Intel FPGA AI Suite DLA Graph

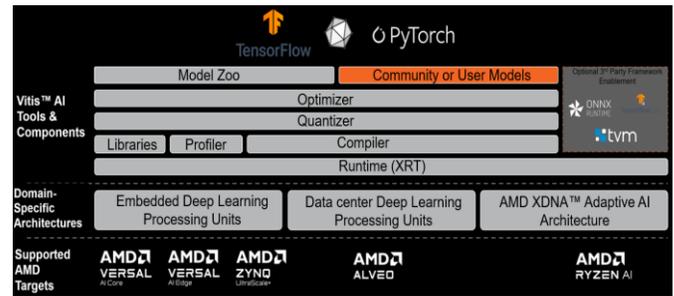

Fig. 12. Vitis AI stack [82].

Compiler then maps network layers to the FPGA, generating hardware instructions, weights, and activations, guided by the .arch IP file that defines the inference.

The toolchain supports ONNX models and applies compiler-level optimizations such as operator fusion, layer reordering, and scheduling, providing a familiar flow for deep learning developers. OpenVINO also enables CPU-FPGA pipeline partitioning for hybrid edge or cloud systems. However, the Intel flow is comparatively less transparent in terms of low-level tunability and hardware mapping. The FPGA AI Suite also lacks a model zoo as extensive as Vitis AI's.

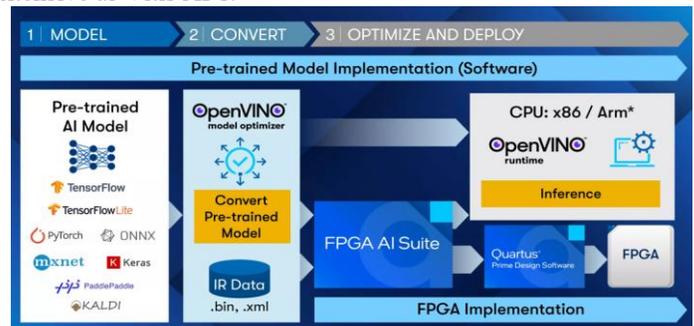

Fig. 13. FPGA AI Inference Development Flow [84].

### C. Other Software Tools

AMD's FINN [85] and python package hls4ml [86] target lightweight and low-precision FPGA deployments using custom HLS and dataflow architectures. FINN starts from a Brevitas-trained QNN (ONNX), applies graph transformations (e.g., folding, quantization), and generates layer-wise HLS C++ kernels via finn-hlslib. These kernels form deeply pipelined 2D PE arrays with minimal off-chip memory access, achieving sub-µs latency even at 1-bit precision. Deployment is automated via Docker or direct Vitis/Vivado projects. hls4ml similarly translates Keras, PyTorch, or ONNX models into synthesizable HLS, producing Vivado/Vitis or Quartus projects with IP cores ready for integration. It offers automatic precision inference, FIFO tuning, and latency-resource trade-off optimization, significantly reducing development time for real-time applications.

Several recent academic efforts have demonstrated substantial performance gains on Xilinx toolchains without modifying the proprietary software itself. AutoHLS [87] improves Vitis HLS via a machine-learning-assisted front end. It samples C/C++ kernels with different pragmas and operations, predicts resource/timing feasibility using DNN/QNN models, and applies Bayesian optimization to converge up to 70× faster than manual exploration,

## VI. CNN Models for Object Detection, classification, and Tracking at the Edge

With the integration of CNNs, vision tasks have grown more computationally intensive, and traditional compute platforms often struggle to meet the stringent energy and latency demands of real-time applications. FPGAs provide a compelling alternative, offering fine-grained, low-level optimizations and the ability to construct custom dataflows tailored to specific model architectures. This section highlights CNN implementations on FPGAs for real-time vision tasks.

### A. Real-time Object Detection on FPGA

Object detection is a fundamental task in computer vision that involves identifying and localizing instances of semantic objects, such as people, vehicles, and animals, within static images or video frames. The advent of Deep Neural Networks (DNNs) enabled the conception of end-to-end learning architecture such that object class labels and locations of bounding boxes are jointly learned from input images with no intermediate modules for feature engineering [88].

CNN-based detectors are typically classified as either one-stage or two-stage. Two-stage detectors such as Region-based Convolutional Neural network (R-CNN) [89], Fast R-CNN [90], and Mask R-CNN [91], first generate candidate regions, then classify them and refine their bounding boxes. While they share core computations (convolutions and matrix multiplications) with one-stage models, they add Region Proposal Networks (RPNs) and ROI alignment, causing irregular memory access patterns, which hinder the parallelism and streaming efficiency that FPGAs are optimized for. However, they yield better accuracy than single-stage counterparts, particularly for small or overlapping objects.

One-stage detectors such as YOLO (You Only Look Once) [92] and SSD (Single Shot MultiBox Detector) [93], perform joint localization and classification in a single step, bypassing region proposals by dividing the image into grids and directly predicting bounding boxes and class labels. YOLO formulates detection as a regression problem, with YOLOv1 pioneering this approach and later versions (YOLOv6–YOLOv11) improving accuracy and efficiency through deeper feature aggregation, spatial attention, and transformer-inspired modules. However, these enhancements increase architectural complexity, making full-scale models less suitable for real-time deployment on resource-constrained FPGAs.

Tiny versions of YOLO are preferred since they significantly reduce the backbone depth, filters, and feature map sizes. For example, Tiny YOLOv3 uses 9 vs. 53 convolution layers, lowers peak channels from 1024 to 512, and reduces MACs by 4–8× [94]. They simplify or eliminate complicated modules such as attention blocks, spatial pyramid pooling, and additional detection heads; They typically predict at two scales instead of three, minimizing parameter count and intermediate feature size.

YOLOv11-Nano and YOLOv11-Small (YOLOv11-S) are two other lightweight variants of the full YOLOv11 model, which were successfully implemented into an FPGA recently, as shown in Fig. 15 [95]. These models demonstrate how carefully designed CNN architecture can overcome the limitations imposed by hardware resource constraints. The hardware accelerator executes YOLOv11 layers sequentially, with routing layers preconfigured with specific memory addresses to improve data access, while loop tiling and burst-mode memory access reduce latency and enhance memory bandwidth. To further boost throughput, multiple parallel PEs operate concurrently on different output channels, aided by data scatter/gather modules and pixel buffers that optimize memory transfers and spatial operations.

YOLOv11-Nano reduces computational complexity using standard and depthwise-separable convolutions, as outlined in Table II. Standard convolutions follow the $K^2 \cdot C_{in} \cdot C_{out} \cdot H_{out} \cdot W_{out}$ compute model, while depthwise-separable convolutions apply 3×3 depthwise + 1×1 pointwise kernels in in two stages to minimize MAC operations. The architecture incorporates C3k2 block (Cross Stage Partial with kernel size 2) for fast processing, and SPPF (Spatial Pyramid Pooling - Fast) and C2PSA (Cross Stage Partial with Spatial Attention) modules to enhance small-object and occlusion detection, as shown in Fig. 14. Cross Stage Partial (CSP) modules improve gradient flow and reduce redundancy without hardware-intensive attention mechanisms [96]. In comparison, YOLOv11-S scales up Nano's design by using deeper CSP modules and wider channels, achieving higher accuracy but at greater resource cost, consuming ≈70% LUTs and 80% DSPs on PYNQ-Z1. It sustains 18 FPS, demonstrating feasible real-time FPGA performance, though the high resource footprint limits scalability for more demanding applications, motivating further FPGA-oriented CNN optimizations [95].

Further improvements in performance and memory efficiency can be achieved by modifying the convolution layers through convolution lowering techniques. In models like YOLOv6-Nano, this strategy transforms conventional k×k convolutions into 1×1 convolutions with more input channels, effectively increasing PE utilization. For example, converting a 3×352×352 image into a 27×176×176 tensor and padding it to 32×176×176 boosts the PE array utilization from just 18.75% to 84.375% [97]. This also reduces DRAM access by nearly 50% compared to naïvely padding the image to match PE dimensions.

CNN models are deployed on DPUs for acceleration. For instance, to accelerate LCAM-YOLOX, which is a YOLOX model with layerwise channel attention module, Elhence et al. [98] utilized DPUCZDX8G and Vitis AI software. LCAM-YOLOX model trained with images of size 640×640 on the SoC-based ZCU102 FPGA achieved real-time performance at 195 FPS, while utilizing only 19.03% of LUTs and 28.17% of DSPs. For detection of airplanes, Brown et.al [99] utilized DPUCVDX8G for YOLOV5 deployment on Versal VCK190 board with a replacement of SiLU layers with LeakyReLU as the DPU does not support SiLU layers. On Rareplane dataset with image size 640×640, this model achieved 244.49FPS and with 0.852 F1 score.

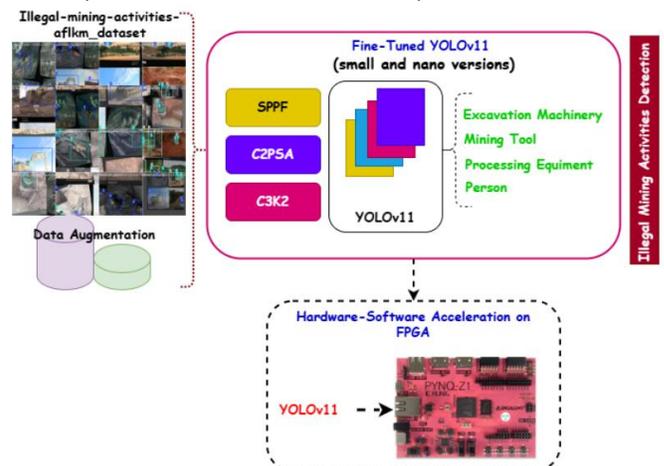

Fig. 14: Implementation of YOLOV11n and YOLOV11s models on PYNQ Z1 FPGA trained for images of size 640×640 [95].

El-Ghany et al. [100] proposed a masked face detection system using Tiny DarkNet, a lightweight CNN optimized for FPGA deployment, using LUT-DSP based Virtex UltraScale+ XCVU19P FPGA. The 21-layer network includes 3×3 and 1×1 convolutions with ReLU activations and is quantized to 16-bit fixed-point for reduced memory and power usage. The Verilog-based accelerator exploits spatial parallelism with 128 convolution and pooling windows. The system achieves real-time inference with 8.3 ms latency, 120.48 FPS throughput, and consumes 4.756 W, utilizing 1153 DSPs and 12% of the available LUTs.

YOLO models, especially tiny and nano variants employ a single feature extraction path and generate dense predictions in a fully convolutional manner. This design aligns well with the spatial pipelining and parallel processing capabilities of FPGAs, enabling more predictable memory access and better resource reuse. Moreover, the regression-based formulation of YOLO avoids the need for large numbers of pre-defined anchors, simplifying control logic and lowering the demand on BRAM and LUT usage.

Another well-known one-stage detector is SSD. SSD predicts class probabilities and bounding box offsets directly from intermediate feature maps using default anchor boxes across multiple scales and aspect ratios, avoiding the region proposal and refinement stages of two-stage models [101]. Multi-scale feature extraction is achieved via backbones such as VGG16 or ResNet50, but this increases memory access, introduces non-uniform computation during anchor box matching and decoding, and complicates FPGA streaming. Deeper backbones with wide channels and parallel multi-feature-map processing further raise DSP and BRAM usage, causing frequent off-chip memory access and reduced pipelining efficiency. Consequently, dynamic power rises significantly; for example, FPGA-based SSD-VGG16 can exceed 80 W that is ≈49× higher than YOLOv3-Tiny, due to heavy DSP usage and wider activations requiring additional buffering and control logic [102], [103].

From the graph in Fig. 15, YOLO models consume less resources, whereas SSD implementations often require larger buffer memory and pipeline balancing but offers higher throughput. Furthermore, FPGA deployments of SSD are less frequently reported with real-time performance, suggesting YOLO's architectural simplicity and efficient convolutional design make it a more favorable candidate for low-latency, high-throughput inference on FPGAs. Further comparisons on recent CNN models are shown in Table VII against their image size, evaluation metrices and performance metrices. From the table, it is evident that the choice of FPGA depends on the application. For instance, if power consumption is a concern, SOC-based Xilinx Zynq+ ZCU106 FPGA consumed less power while implementing the same YOLOV3-tiny on LUT-DSP based Xilinx XCKU060 FPGA consumed high power of 8.1W and high DSPs but low LUTs. Resource usage and power consumption of SoC based FPGAs and those with hardcore AI engines can be compared from the table for YOLOV2 with tiny darknet. Versal Premium XCVP1902 consumed less power ad they utilized only 4% of LUTs, 13% BRAMs, 17% DSPs and 1% FFs.  If speed is the main criteria, LUT-DSP based Xilinx XCKU060 FPGA comes to the top.

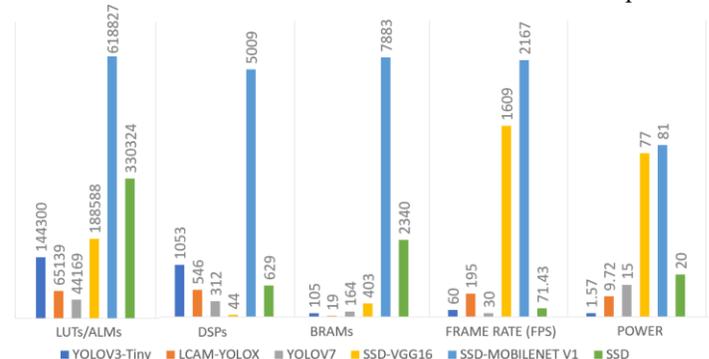

Fig. 15: Graphical comparison of YOLO and SSD models based on their resource utilization and throughput.

TABLE VII
COMPARISON BETWEEN RECENT REAL-TIME DETECTION METHODS ON FPGA

| Ref. | Detection Model | FPGA | Freq. (MHz) | Frame Rate (FPS) | mAP % | Image Size | Latency (ms) | Through-put (GOPS) | Power (W) | BRAM | LUTs/ALMs | DSP | FFs |
|---|---|---|---|---|---|---|---|---|---|---|---|---|---|
| [103] | YOLOv3-tiny | Xilinx Zynq+ ZCU106 | 125 | 60 | - | 320×320 | 5.13 | 292.18 | 1.57 | 105 | 144.3k | 1053 | 200.1k |
| [98] | LCAM-YOLOX | Kria KV260 | 300 | 195 | 84.73 | 640×640 | 11.28 | - | 9.72 | 19 | 65139 | 546 | 108532 |
| [104] | SRNET | XCZU15EG | 200 | 40.5 | 86.68 | - | 24.7 | 150 | 7.84 | - | 47012 | 1592 | - |
| [105] | YOLOv3-Tiny | Xilinx XCKU060 | 200 | 62.8 | - | 416×416 | - | 408 | 8.1 | 1485 | 43.2k | 1364 | 94.4k |
| [100] | YOLO-v2 with Tiny Darknet | UltraScale+ XCVU19P | 80 | 120.48 | - | 226×226 | 8.3 | - | 6.014 | 928.5 | 484209 | 1153 | 156665 |
| [100] | YOLO-v2 with Tiny Darknet | Versal Premium Series XCVP1902 | 80 | 120.48 | - | 226×226 | 8.3 | - | 4.756 | 892.5 | 359591 | 1153 | 123848 |
| [102] | Custom SSD | ZCU 102 | 350 | 1609 | 25.1 | 300×300 | - | - | 77 | 403 | 188588 | 44 | - |

## B. Real-time Object Classification on FPGA

Real-time object classification refers to the task of assigning a semantic label to an object or region within an image or video frame. In deep learning pipelines, CNNs extract features that are passed to fully connected layers for class probability estimation via softmax. Models such as MobileNet [106], ResNet [107], and EfficientNet [108] are widely used for video classification due to their efficiency–accuracy balance and map well to FPGA accelerators by exploiting spatial locality for real-time inference.

The recent classification model, MobileNetV4 (MNv4), represents a significant architectural leap, introducing a new building block called the Universal Inverted Bottleneck (UIB) and an accelerator-optimized attention unit termed Mobile MQA, as illustrated in Fig. 16 [39]. UIB serves as a superset of lightweight convolutional blocks, supporting Inverted Bottlenecks (as in MobileNetV2), ConvNeXt-like modules, ExtraDW configurations, and FFN-style blocks, with optional depthwise and pointwise/2D convolutions for adaptive spatial and channel processing. A fused IB variant simplifies hardware mapping, while two optional DW layers enable resolution control.

Additionally, MNv4 introduces a refined Neural Architecture Search (NAS) and multi-dataset distillation to achieve Pareto-optimal performance across CPUs, DSPs, GPUs, and edge accelerators like Google's EdgeTPU [109] and Apple's Neural Engine [110]. In particular, Mobile MQA, while highly efficient on fixed-function EdgeTPUs, is difficult to replicate efficiently on FPGAs. Although systolic-like dataflows can be implemented, reproducing the fine-grained spatial reduction and Einsum-style matrix contractions demands considerable logic and routing resources. This results in suboptimal hardware utilization, increased latency, and bandwidth bottlenecks, especially under resource-constrained FPGA environments.

Additionally, UIB's flexible structure with optional depthwise layers introduces architectural irregularity, leading to routing congestion and inefficient mapping, unlike the FPGA-friendly MobileNetV1–V3. Finally, NAS and distillation-based training, while improving accuracy, offers no runtime benefit unless explicitly re-targeted for FPGA synthesis. Consequently, EfficientNet and ResNet remain the preferred models for FPGA classification due to their regular, hardware-aligned structure.

EfficientNetV2 [111] introduces progressive learning and a hybrid design combining MBConv (depthwise separable with SE) and Fused-MBConv (standard convolutions replacing early depthwise layers) to simplify memory access and improve hardware efficiency. This design balances latency, arithmetic intensity, and resource use across embedded platforms. However, SE modules and Swish activations are compute-intensive, so FPGA deployments often approximate or replace them with ReLU or Hard-Swish to reduce complexity while preserving accuracy.

ResNet [112], though less parameter-efficient, is widely deployed on FPGAs due to its modular residual blocks, uniform tensor shapes, and repetitive computation patterns that map efficiently to parallel FPGA data paths. These characteristics make variants such as ResNet-18 and ResNet-50 common baselines for benchmarking and prototyping. For instance, Fu et al. [113] proposed a real-time railway fault classification system using Binary Neural Networks (BNNs) derived from ResNet (RBPnet2 and RBPnet4), deployed on Zynq-based edge platforms. As shown in Fig. 17, the system is divided into an image acquisition and control stage, an accelerator computing engine (handling convolution, batch normalization, and activation), and external memory interaction for weights and feature maps. The classification result is post-processed using average pooling and softmax operations. A FIXED12 DSP-packing scheme replaced multiply-accumulate operations with XNOR–popcount logic, improving resource efficiency and reducing LUT usage by 10%.

Although classification models assign a single label to an entire image, their intermediate convolutional layers generate spatial feature maps that preserve local information, making them effective feature extractors in detection frameworks. Detectors such as YOLO and SSD use classification networks (e.g., ResNet, MobileNet, EfficientNet) as backbones for hierarchical feature extraction, discarding the final classification layer, while performing joint localization and classification. Therefore, they are not used for standalone image classification tasks where only a single class label per image is needed, as that would be unnecessarily complex and inefficient. Unlike detection, classification models apply softmax or exponential functions to convert raw scores to probabilities, requiring normalization across the output vector. These operations are more computationally intensive than basic convolution or activation layers and must be carefully optimized in real-time FPGA deployments to prevent latency bottlenecks.

Fig. 18a compares different platforms based on accuracy, FPS and GMAC of a lightweight attention-based CNN for drone surveillance [114]. This network interleaves five convolution blocks, each composed of four standard and one depth-wise separable convolution followed by ReLU and max-pooling, with four attention blocks that use expanded receptive fields to recalibrate spatial feature maps. The model achieved 947.2 FPS, 53.8 GMAC, and slightly lower accuracy than the pure PS implementation, with the ZCU104 FPGA outperforming both PYNQ-Z2 and Google TPU with Raspberry Pi 4. Among the FPGAs, ZCU104 also showed lower resource utilization as shown in Fig. 18b.

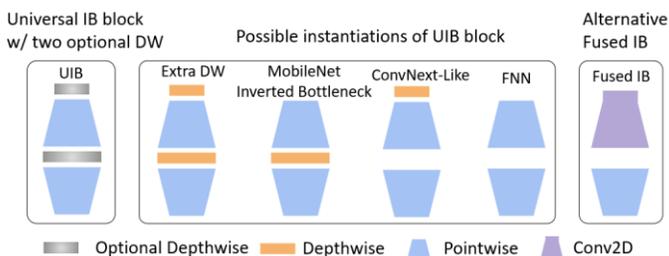

Fig.16. The UIB block in MobileNetV4 is a configurable module that generalizes common CNN blocks [39].

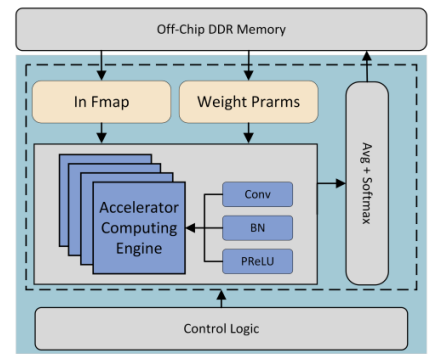

Fig. 17: FPGA-based BNN accelerator proposed in [113], showing interactions between the control logic, accelerator engine, DDR, and post-processing modules.

Table VIII compares few recent CNN based classification models on FPGA. While models like U-Net and LightLiteNet emphasize accuracy, others like RBPnet4 are optimized for speed and efficiency on low-power FPGAs. RESNET-14 reveal the challenge of scaling throughput without incurring significant latency on LUT-DSP based FPGA architectures. RBPnet4, targeting Zynq 7020, delivers the best latency (0.7 ms) and a competitive 60 FPS frame rate while consuming 2.3 W, making it highly efficient for real-time embedded scenarios. Therefore, SoC based FPGAs are used for CNN based classification tasks.

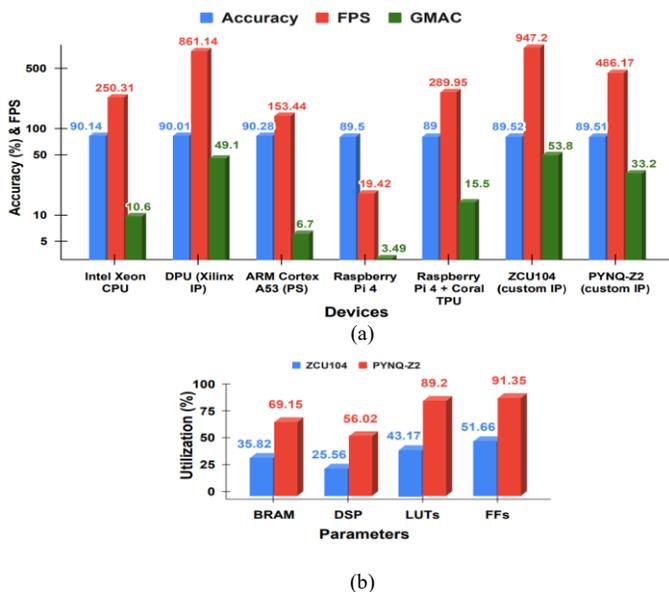

Fig. 18: (a) Comparison of performance matrices across different platforms (b) Resource utilization for ZCU104 and PYNQ-Z2 [114].

### C. Real-time Object Tracking on FPGA

Real-time object tracking focuses on detecting and following objects in video streams with minimal latency, which is essential for autonomous vehicles, robotics, surveillance, and AR applications[115], [116]. Tracking in crowded or dynamic scenes enables early anomaly detection in surveillance and predictive navigation for vehicles [116]. In addition, tracking is required for vehicle navigation as it predicts the future position of eventual objects in the scene.

Traditional correlation- and Kalman-filter trackers are simple but fail under occlusion or appearance changes [117]. CNN-based trackers overcome these issues by leveraging rich spatio-temporal features for robustness to lighting, background, and object variations. Early end-to-end designs like GOTURN regress target bounding boxes from consecutive frames [133] while lightweight trackers such as SiamDW, LightTrack, and OceanLite reduce network depth and operations, sustaining >60 FPS on modern GPUs. With quantization and pruning, these models are adaptable for real-time embedded FPGA deployment, balancing latency and resource efficiency.

There have been also encouraging trade-offs between speed and accuracy in hybrid approaches that combine deep features with classical methods, particularly Siamese network architectures [115], [118]. For instance, the Siamese Fully Convolutional (SiamFC) tracker, illustrated in Fig. 19, learns a similarity metric between a target template and a search region using two identical convolutional branches with shared weights [118]. These branches extract feature maps $\varphi(z)$ and $\varphi(x)$ from the template image $z$ and search image $x$, respectively. The object localization is determined by computing a response map $R(x, z)$ via cross-correlation:

$$R(x, z) = \varphi(z) * \varphi(x) \quad (5)$$

where * denotes the cross-correlation operation between the two feature maps. The peak of $R(x, z)$ gives the most probable object location, enabling one-shot inference without online training. For efficiency, template features $\varphi(z)$ are computed once and cached, while $\varphi(x)$ is computed per frame. The cross-correlation is performed over all positions in the search map, yielding a computational cost that is proportional to the size of the feature maps and number of channels.

FPGA implementations that utilize SiamFC typically adopt lightweight CNNs such as AlexNet [119] for shared feature extraction, employing only the five convolutional layers with interleaved ReLU and max pooling, omitting fully connected layers to reduce memory and latency. Given the fixed input sizes in SiamFC, each forward pass involves a predictable sequence of 2D convolutions and activations that are repeated frame-by-frame. This regular feed-forward structure simplifies deployment and enables significant speedups through hardware-software co-design, yielding throughput close to floating-point baselines on FPGA platforms.

TABLE VIII
COMPARISON BETWEEN RECENT REAL-TIME CLASSIFICATION METHODS ON FPGA

| Ref. | Model | FPGA | Frequency (MHz) | Frame Rate (FPS) | Accuracy Top 1 | Image Size | Latency (ms) | Through put (GOPS) | Power (W) | BRAM | LUTs/ ALMs | DSP | FFs |
|---|---|---|---|---|---|---|---|---|---|---|---|---|---|
| [120] | U-Net | Zynq UltraScale+ MPSoC. | - | 37.5 | 99 | 256×256 | 26.7 | - | 2.4 | - | - | - | - |
| [121] | VGG16 | Avnet Ultra96-V2 | 325 | - | 89.54 | 32×32 | 0.652 | - | - | 126 | 38,418 | 326 | 58,831 |
| [122] | RESNET 14 | Xilinx XCVU9P | 100 | 34.7 | 65.9 | 256×192 | 75.3 | 1083 | 25 | 2035 | 760k | 3982 | - |
| [123] | LightFire Net | ZYNQ Z7-Lite 7020 | - | 15 | 96.70 | 48×48 | - | - | 2.23 | 64 | 39692 | 150 | 49561 |
| [113] | RBPnet4 | Zynq 7Z020 | 100 | 60 | 0.93 (F1 score) | 32×32 | 0.7 | 77.52 | 2.3 | 144 | 65492 | 160 | - |

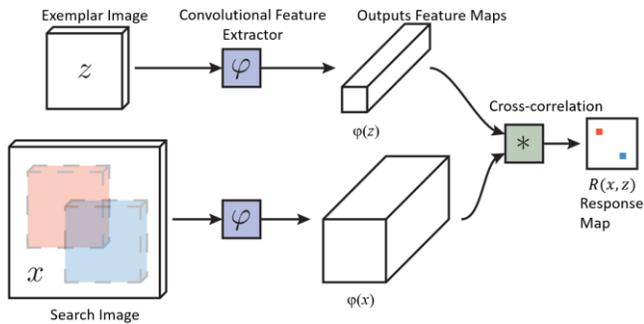

Fig. 19: Fully Convolutional Siamese Architecture obtained from [118]. Similarity scores for the sub-windows are shown in red and blue pixels of score map.

In contrast to end-to-end trackers, DeepSORT [124] employs a tracking-by-detection approach, using a pretrained CNN for object detection and a secondary CNN for appearance embeddings, combined with Kalman filter-based motion estimates for data association. Though not a pure end-to-end CNN tracker, DeepSORT's CNN-based feature extraction supports robust MOT. FPGA platforms for MOT often adopt this tracking-by-detection approach to balance flexibility and performance. Detection is offloaded to PL, while object association and control logic execute on the PS, enabling parallelism and low-latency inference.

A notable example is the real-time MOT system by Danilowicz and Kryjak [125], implemented on the Xilinx Zynq UltraScale+ MPSoC. It uses a 4-bit quantized YOLOv8n detector accelerated in PL (via the FINN framework) and the SORT algorithm on the ARM PS, as illustrated in Fig. 20. The detector employs pipelined dataflow modules with quantized convolution, activation, and thresholding IP cores, optimized using Brevitas/QONNX transformations. After detecting bounding boxes in PL, the PS handles non-maximum suppression and data association via SORT. Efficient communication between PL and PS is maintained using deep FIFOs and DMA, sustaining real-time throughput. The system achieved 195.3 FPS for detection alone and 24 FPS end-to-end due to Python-based postprocessing. It attained 0.389 MOTA on MOT15 and competitive scores on UA-DETRAC benchmarks. The compact 12 MB detector achieved mAP comparable to full-precision models while consuming 38% LUTs and 13% FFs on the ZCU102 FPGA. These results demonstrate a strong balance of accuracy, hardware efficiency, and speed, making the design well-suited for embedded MOT applications.

Table IX summarizes recent FPGA-based tracking systems, each optimized for different trade-offs. From the table, it is evident that for tracking tasks, SoC-based FPGAs are preferred, as no literatures focused on other two types possibly due to the need for tight integration of control-intensive algorithms (e.g., Kalman filtering, data association) with real-time operating systems and high-level I/O stacks. SiamRPN++ [93] on Kria KV260 prioritizes low power (1.11 W) but achieves only 6.37 FPS. In contrast, YOLOv3-tiny with DeepSORT [168] on Zynq-7000 achieves 168.72 FPS and 67.91 GOPS throughput, but at higher power (15.18 W), reflecting a speed-optimized, pipelined design.

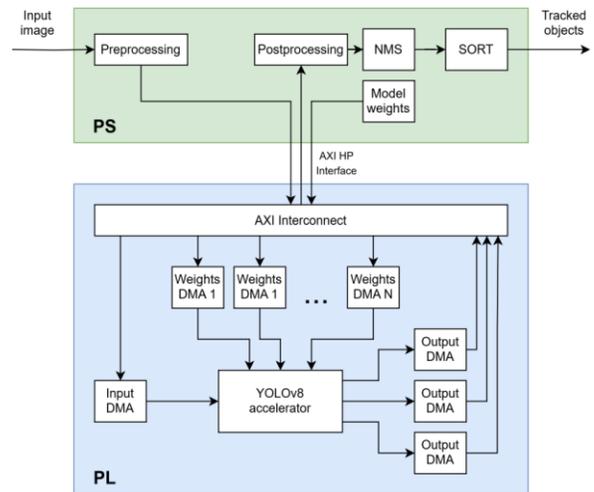

Fig. 20: Task partitioning in Zynq UltraScale+ MPSoC MOT: PL runs 4-bit pipelined YOLOv8n inference; PS handles NMS and SORT tracking via DMA/FIFOs [125].

YOLOv8n + SORT [56], running on Zynq UltraScale+, combines quantized detection in hardware with tracking in software, achieving 24 FPS and 0.389 MOTA while maintaining efficient resource use. Lastly, the CNN + BG-Diff tracker [169] balances speed (54.67 FPS) and low power (5.5 W) with minimal hardware usage, making it ideal for lightweight applications. These implementations collectively showcase that the choice of tracking model and FPGA platform must align with specific design goals, whether it is maximizing throughput, minimizing latency and power, or balancing accuracy for multi-object scenarios. Lightweight CNNs and quantized YOLO variants emerge as favorable candidates for high-throughput designs, whereas Siamese networks and hybrid trackers offer modularity and lower power footprints, benefiting edge deployments.

### D. Heterogenous Systems

Utilizing the advantages of each processor type, heterogeneous systems that combine CPUs, Neural Processing Units (NPUs), GPUs, and FPGAs provide an intriguing architecture for real-time vision and AI inference. Microsoft's Project Brainwave [126] realizes a soft NPU on Intel FPGAs, optimized for real-time AI inference in cloud deployments. The architecture, as illustrated in Fig. 21, centers around a custom Matrix Vector Unit (MVU) synthesized directly into FPGA logic, delivering low-latency, batch-1 inference by avoiding host CPU and DRAM bottlenecks [127]. Model weights are persistently stored in on-chip SRAM and tightly coupled memory, ensuring consistent performance A single-threaded instruction controller issues wide, parallel operations across thousands of MAC units, orchestrated through a custom SIMD-like ISA and hardware scheduler. This design supports major CNN workloads and is reconfigurable, allowing rapid deployment of evolving models without silicon changes, highlighting the flexibility and scalability of FPGA-based NPU implementations in real-time object detection and classification systems.

TABLE IX
Comparison Between Recent Real-time Tracking Methods on FPGA

| Ref. | Tracking Model | FPGA | Freq. (MHz) | Frame Rate (FPS) | Accuracy (%) | Image Size | Latency (ms) | Through put (GOPS) | Power (W) | BRAM | LUTs/ ALMs | DSP | FFs |
|---|---|---|---|---|---|---|---|---|---|---|---|---|---|
| [128] | SiamRPN++ | Kria KV260 | - | 6.37 | 56.3 | - | 157.12 | - | 1.11 | 91 | 70006 | 765 | 128663 |
| [129] | YOLOv3-tiny + Deepsort | Zynq-7000 | 209 | 168.72 | 59.9 | 416×416 | | 67.91 | 15.18 | 263 | 91108 | 294 | 89148 |
| [125] | YOLOV8n + modified SORT | Zynq UltraScale+ | 300 | 24 | 38.9 MOTA | - | - | - | - | 421 | 205k | 486 | 246k |
| [130] | CNN + BG-Diff Tracker | Zynq UltraScale XCZU9EG | 220 | 54.67 | | 227×227 | - | - | 5.5 | 223 | 48.5k | 220 | - |

A heterogeneous CPU/FPGA/NPU architecture, shown in Fig.22, has been proposed for accelerating UAV vision tasks such as classification (MobileNet), detection (YOLO), FFT, and filtering [131]. A task-aware scheduler dynamically assigns workloads based on latency and energy profiles: the FPGA handles data-parallel filtering and signal processing via streaming and reconfiguration, the NPU accelerates deep models using tensor cores, and the CPU manages control and light tasks. The system supports automated scheduling, dynamic task-to-core mapping, and partial FPGA reconfiguration, with KubeEdge enabling lightweight edge-cloud orchestration.

Experimental results across CPU, CPU/GPU, CPU/FPGA, CPU/NPU, and CPU/FPGA/NPU variants highlight the efficiency gains. Bilateral filtering drops from 411.65 ms/3.44 J on the CPU to 58.27 ms/1.29 J on the FPGA and 36.27ms/1.21J on the full system. FFT reduces from 123.42 ms/1.86 J on CPU to 12.89 ms/0.48 J on FPGA and 10.21 ms/0.43 J on the heterogeneous system. YOLOv5s inference (650×433) decreases from 2783.32 ms/6.33 J on CPU to 35.72 ms/1.04 J on the NPU. Overall, the CPU/FPGA/NPU architecture achieves up to 87% energy savings and >70× speedup over CPU-only execution.

The ability of CPU-FPGA and GPU-FPGA platforms to balance throughput, latency, and power efficiency has led to their increasing popularity in edge vision and language processing. By speeding up memory-bound operations and facilitating dataflow optimizations, FPGAs enhance GPUs, which provide high FLOP rates appropriate.

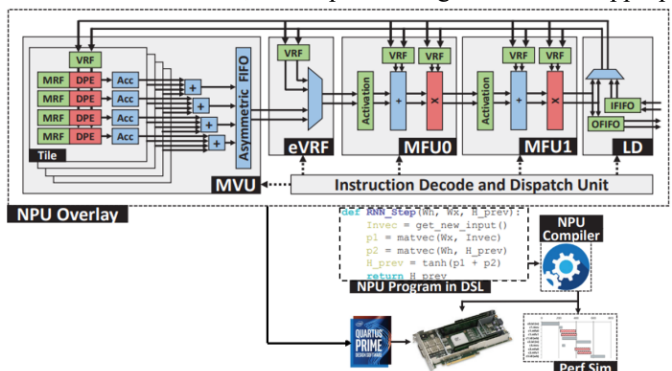

Fig. 21: Overview of the NPU overlay architecture [127].

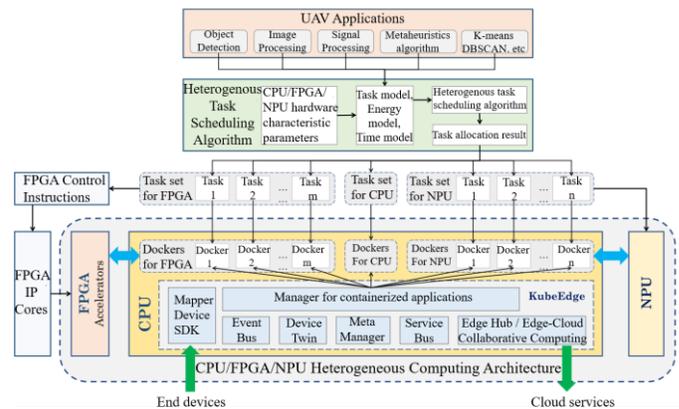

Fig. 22: CPU/FPGA/NPU heterogeneous architecture for UAV vision tasks, showing task-specific accelerator mapping and system coordination via the CPU [131].

for dense tensor operations. For these systems to fully utilize resource synergy, intelligent runtime schedulers and tightly coupled interconnects (such as PCIe, AXI, and CCIX) are necessary. These works illustrate how important it is to divide tasks according to precision, parallelism, and access patterns to maximize the benefits of heterogeneity in vision inference

## VII. Design Trade-Offs and Challenges in FPGA-Based Vision Inference

Deploying deep learning models on FPGAs for real-time detection, classification, and tracking requires managing interdependent trade-offs in model architecture, memory hierarchy, compute efficiency, and resource utilization. These challenges are amplified when complex CNNs—with residual connections, multi-scale fusion, and attention modules—demand trillions of operations per second. FPGAs, constrained by limited on-chip memory and fixed interconnects, rely on careful tiling and pipelined dataflows, where activations and weights are partitioned into BRAM-fit tiles. Yet, tile dimensions vary by layer, and over-tailored tiling can introduce control overhead or leave compute units underutilized when layer dimensions are irregular.

Precision scaling alleviates bandwidth and resource pressure by reducing FP32 weights and activations to INT8, INT4, or even binary formats, enabling multiple MACs per DSP slice and improving throughput per watt. Yet aggressive quantization risks accuracy loss in sensitive layers such as those with normalization or attention, while non-uniform or mixed-precision schemes increase LUT usage and timing closure complexity.

To run CNNs efficiently on FPGAs, designers need to decide how to divide up the work. One option is to break the computations into many small parts and run them in parallel. This can make the system very fast, but it also uses a lot of FPGA resources and can be difficult to route signals between parts. Another option is to group the work into larger blocks and process them in stages. This approach is easier to manage and uses fewer resources, but it might not fully use all the parallel hardware, especially for large layers. Because finding the best setup is complex, automated tools are often used to explore different designs and choose the most efficient one.

The software ecosystem further constrains practical deployment. HLS flows such as Vitis AI, FINN, hls4ml, and OpenVINO accelerate model-to-bitstream workflows but may underutilize hardware compared to hand-crafted RTL, which achieves maximum efficiency at the cost of longer development and verification cycles. Moreover, Xilinx's regular updates to Vitis AI (e.g., 2025.03) have added new Layer-Fold optimizations and DPU scheduling enhancements that directly accelerate Vision Transformer and YOLO workloads on Alveo and Versal devices. Whereas comparable Intel FPGA optimizations in OpenVINO 2024.2 primarily target CPU inference and only secondarily improve FPGA paths.

Future FPGA development for deep learning is expected to become smarter, more flexible, and more accessible. Current LUT-DSP FPGAs and ACAPs with AI Engines require advances in compiler and runtime frameworks to efficiently handle sparse and irregular dataflows, manage on-chip memory hierarchies, and schedule control-intensive tracking kernels alongside high-throughput CNN pipelines. Emerging intelligent compilers will automatically map CNN models to FPGA resources, reducing manual design effort. Partial reconfiguration may enable loading different accelerators for distinct model stages, while built-in support for attention, sparse computation, and low-precision arithmetic will improve the efficiency of Transformers and CNN-ViT hybrids. Finally, full-stack toolchains integrating training, quantization, hardware mapping, and edge deployment will make FPGA adoption more practical for real-time vision applications.

## VII. Conclusion

This paper has surveyed the evolving landscape of FPGA-based acceleration for CNN-driven object detection, classification, and tracking. FPGAs offer distinct advantages for real-time, power-efficient vision tasks at the edge, thanks to their reconfigurability, custom dataflows, and support for low-precision arithmetic. Through reviewing hardware-aware model optimizations, parallelism strategies, and software toolchains, we highlighted the trade-offs between performance, resource utilization, and model accuracy. Additionally, hybrid architectures combining CPUs, GPUs, and FPGAs present promising avenues for scaling AI inference in heterogeneous systems. As deep learning models grow in complexity, future work will benefit from dynamic pipeline optimizations, improved sparsity exploitation, and integrated compiler frameworks that simplify design space exploration. These advancements will be crucial in harnessing FPGAs for next-generation real-time vision applications across diverse and resource-constrained environments.